\documentclass[letterpaper,11pt]{article}
\pdfoutput=1

\usepackage{jheppub}
\usepackage{mathtools,amssymb,braket,mathrsfs,tikz,subfig,xspace,xcolor,float,color,siunitx,comment,afterpage,multirow,array,hhline,amsmath,amsthm,framed,colortbl, bm}
\usepackage[countmax]{subfloat}

\newcommand{\be}{\begin{equation}}
\newcommand{\ee}{\end{equation}}

\newcommand{\E}{\mathcal{E}}

\newcommand{\EMD}{\text{EMD}}
\newcommand{\SEMD}{\text{SEMD}}

\newcommand{\tot}{\text{tot}}

\renewcommand{\O}{\mathcal{O}}

\usepackage{amsmath,amssymb,amsthm,cancel,hyperref,graphicx,xcolor}
\usepackage{mathrsfs,wasysym}
\usepackage{booktabs}
\usepackage{tikz}
\usetikzlibrary{calc}
\usepackage{placeins}

\usepackage{slashed}
\usepackage{graphicx}

\usepackage{soul}

\usepackage{array}

\usepackage{comment}

\usepackage{url}
\usepackage{xspace}
\usepackage{dsfont}
\usepackage{amssymb}
\usepackage{amsmath}
\usepackage{graphicx}
\usepackage{hhline}
\usepackage{physics}

\usepackage{color}
\definecolor{mit-red}{rgb}{0.64,.12,0.2}
\definecolor{darkred}{rgb}{1.0,0.1,0.1}
\definecolor{darkgreen}{rgb}{0.1,0.7,0.1}
\definecolor{darkblue}{rgb}{0.1,0.1,1.0}

\usepackage{amsmath}

\DeclareMathOperator*{\argmin}{argmin}

\DeclareRobustCommand{\Sec}[1]{Sec.~\ref{sec:#1}}
\DeclareRobustCommand{\Secs}[2]{Secs.~\ref{sec:#1} and \ref{sec:#2}}
\DeclareRobustCommand{\App}[1]{App.~\ref{app:#1}}

\DeclareRobustCommand{\Fig}[1]{Fig.~\ref{fig:#1}}
\DeclareRobustCommand{\Figs}[2]{Figs.~\ref{fig:#1} and \ref{fig:#2}}
\DeclareRobustCommand{\Figss}[3]{Figs.~\ref{fig:#1}, \ref{fig:#2}, and \ref{fig:#3}}

\DeclareRobustCommand{\Eq}[1]{Eq.~(\ref{eq:#1})}
\DeclareRobustCommand{\Eqs}[2]{Eqs.~(\ref{eq:#1}) and (\ref{eq:#2})}
\DeclareRobustCommand{\Reference}[1]{Ref.~\cite{#1}}
\DeclareRobustCommand{\References}[1]{Refs.~\cite{#1}}

\newcommand{\FastJet}{{\sc FastJet}\xspace}

\newcommand{\Shaper}{\textsc{Shaper}\xspace}
\newcommand{\Specter}{\textsc{Specter}\xspace}

\begin{document}

\count\footins = 1000
\interfootnotelinepenalty=10000
\setlength{\footnotesep}{0.6\baselineskip}

\renewcommand{\arraystretch}{1.3}

\title{SPECTER: Efficient Evaluation of the Spectral EMD}

\author[a,b]{Rikab Gambhir,}
\author[c]{Andrew J. Larkoski}
\author[a,b]{and Jesse Thaler}

\affiliation[a]{Center for Theoretical Physics, Massachusetts Institute of Technology,  Cambridge, MA 02139, USA}
\affiliation[b]{The NSF AI Institute for Artificial Intelligence and Fundamental Interactions, USA}
\affiliation[c]{Department of Physics and Astronomy, University of California, Los Angeles, CA 90095, USA\\
Mani L. Bhaumik Institute for Theoretical Physics, University of California, Los Angeles, CA 90095, USA}
\emailAdd{rikab@mit.edu}
\emailAdd{larkoa@gmail.com}
\emailAdd{jthaler@mit.edu}

\preprint{MIT-CTP 5771}

\abstract{The Energy Mover's Distance (EMD) has seen use in collider physics as a metric between events and as a geometric method of defining infrared and collinear safe observables.
Recently, the Spectral Energy Mover's Distance (SEMD) has been proposed as a more analytically tractable alternative to the EMD.
In this work, we obtain a closed-form expression for the Riemannian-like $p = 2$ SEMD metric between events, eliminating the need to numerically solve an optimal transport problem.
Additionally, we show how the SEMD can be used to define event and jet shape observables by minimizing the distance between events and parameterized energy flows (similar to the EMD), and we obtain closed-form expressions for several of these observables.
We also present the \Specter framework, an efficient and highly parallelized implementation of the SEMD metric and SEMD-derived shape observables as an analogue of the previously-introduced \Shaper for EMD-based computations.
We demonstrate that computing the SEMD with \Specter can be up to a thousand times faster than computing the EMD with standard optimal transport libraries.
}

\maketitle

\section{Introduction}

Events from particle physics collisions live in a space rich with structure, whose dimensionality, topology, curvature, and other geometric properties encode the myriad physical phenomena that can be observed.
A first step to quantifying the features of this space is to define a metric distance between events, and further ideally in an infrared and collinear (IRC) safe way to render the parametrization of the space calculable in perturbative quantum field theory.
The first IRC-safe metric on the space of collider events was constructed in \Reference{Komiske:2019fks} and called the Energy Mover's Distance (EMD), which quantified the minimal energy cost of deforming one event into another.  
The EMD has since sparked significant interest, with numerous other metrics proposed or exploiting properties of the EMD in various contexts~\cite{Mullin:2019mmh,Larkoski:2020thc,Cai:2020vzx,Cesarotti:2020hwb,Cai:2021hnn,CrispimRomao:2020ejk,Tsan:2021brw,Kitouni:2022qyr,Shenoy:2023ros,Davis:2023lxq,Alipour-fard:2023prj,Gaertner:2023tzv,Craig:2024rlv,Cai:2024xnt}.
This includes the \Shaper framework~\cite{Ba:2023hix}, which can be used to construct and evaluate generalized EMD-based event and jet shapes.
EMD-based observables have also received heightened interest in the experimental community --  one such observable, the event Isotropy~\cite{Cesarotti:2020hwb, Cesarotti:2024tdh}, was recently measured at the Large Hadron Collider (LHC) by the ATLAS collaboration~\cite{ATLAS:2023mny}.

The EMD is formulated as the Wasserstein metric \cite{kantorovich1939mathematical,wasserstein1969markov,dobrushin1970prescribing} of an optimal transport problem in which energy deposits from one event are moved around the detector to reproduce the energy deposits of another event.
An idealized particle physics detector lives on the celestial sphere, so this optimal transport problem is defined in two dimensions.  
Because objects in two or higher dimensions lack a natural ordering, there is no general closed-form expression for the EMD as a function of event constituents.
Thus, generically, this distance must be evaluated through some numerical optimization.
This irreducible numerical computation obscures analytical understanding of the EMD, except in the simplest of event configurations, and further can suffer from slow evaluation, though there are now efficient codes for optimal transport in two or more dimensions \cite{flamary2021pot,Ba:2023hix}.

With these points as motivation, the Spectral EMD (SEMD) was introduced recently~\cite{Larkoski:2023qnv}.  
Instead of optimal transport on the celestial sphere, the SEMD is defined through optimal transport between spectral function representations of events \cite{Basham:1978bw,Tkachov:1995kk,Jankowiak:2011qa,Chen:2020vvp,Chakraborty:2019imr}.
Spectral functions are one-dimensional distributions that encode pairwise particle angles and energy products.  
As such, the spectral representation is automatically invariant to isometries of the events being compared.
Further, a theorem of \Reference{BOUTIN2004709} ensures that the spectral representation can be, with probability 1, mapped to a unique distribution of particles on the celestial sphere, up to isometries.
This result establishes that the SEMD is not only an approximation to the EMD or only approximately a metric, but is indeed a genuine metric between collision events satisfying identity of indiscernibles, symmetry, and the triangle inequality.
Crucially, because points in one dimension are well ordered, optimal transport in one dimension can be expressed as an integral of the difference of the spectral quantile functions (inverse cumulative spectral functions), which can correspondingly be evaluated very efficiently.

Nevertheless, \Reference{Larkoski:2023qnv} mostly focused on the $p=1$ Wasserstein metric of the spectral representation.  
While simple and motivated historically by the Monge-Kantorovich problem \cite{monge1781memoire} and the Earth Mover's Distance from computer vision~\cite{192468, 10.5555/938978.939133, Rubner2004TheEM, Pele2008ALT}, direct evaluation of this integral between arbitrary events is challenging because its integrand is expressed as the absolute value of a difference of functions.
Thus, in practice, the $p=1$ SEMD is, like the ordinary EMD, most easily numerically evaluated.
Further, $p=1$ Wasserstein metrics lack properties that, as physicists, we have come to expect of metrics, namely, their direct interpretability as differential line elements with a natural Riemannian structure.
Simply modifying the definition of the SEMD to a $p=2$ Wasserstein metric endows the SEMD with many nice properties familiar from Riemannian metrics \cite{lott2006ricci,lott2007geometric,villani2008optimal, Cai:2020vzx}.

In this paper, we flesh out the $p=2$ SEMD metric, establish its remarkable properties, and introduce a computational framework called \Specter for its efficient evaluation.
First and foremost, the $p=2$ SEMD can be expressed exactly in closed-form between any two events as represented as point clouds on the celestial sphere.
This represents the \emph{first} metric on collider events that can be expressed with no approximations or need for numerical evaluation, and correspondingly renders theoretical calculations potentially feasible.
Novel event and jet shape observables can be constructed from the SEMD as defined by the minimal distance from an event of interest to a desired energy flow configuration as in \References{Komiske:2020qhg,Ba:2023hix}, and essentially becomes a problem in geometric probability \cite{solomon1978geometric,klain1997introduction,santalo2004integral}, formulated as shape-line picking.
As such, many observables constructed from the SEMD enjoy a closed-form expression as well.
Moreover, we argue that SEMD and derived observables capture the same QCD jet physics as do the EMD and its observables, potentially enabling high-speed precision studies of QCD, while being complementary to the EMD for non-QCD physics.

To facilitate these computations, we construct the \Specter framework, available at \url{https://github.com/rikab/SPECTER}. 
\Specter is to the SEMD what \Shaper is to the EMD: an efficient program for evaluation of the $p=2$ SEMD on events from data (real or simulated), and for the construction and evaluation of \emph{all} spectral event and jet shape observables.%
\footnote{Unlike \Shaper, though, \Specter does not correspond to some tortured acronym.}
Due to the closed-form expression and well-ordering in one dimension, \Specter evaluates the SEMD between events hundreds or even thousands of times faster than comparable codes can evaluate the EMD.
On an ordinary GPU architecture, \Specter can perform 1 million metric evaluations per second, making it far more feasible to compute distances between all pairs of events (to compute correlation dimensions~\cite{Komiske:2022vxg}), perform gradient descent (for parameter optimization~\cite{Ba:2023hix}), or experimental studies (potentially even at the trigger level).

The organization of this paper is as follows.
In \Sec{p2SEMD}, we review the spectral representation of events, evaluate the $p=2$ SEMD, and discuss its properties.
In \Sec{specter}, we introduce the \Specter framework and discuss the algorithm to efficiently evaluate the SEMD between events with $N$ particles in only $\O(N^2 \log N)$ time compared to $\O(N^3)$ for the EMD, and demonstrate performance through some runtime benchmarks.
In \Sec{specshapes}, we review how to construct event and jet shape observables from a metric formulation, and present a collection of observables defined from the SEMD useful for identifying structure on both complete events as well as on individual jets.
In \Sec{compstudies}, we study the SEMD and derivative observables on simulated events, providing comparisons where possible to the EMD.
We conclude in \Sec{concs}, and present some more details of ground metric choices and closed-form expressions in the appendices.

\section{The $p = 2$ Spectral EMD}\label{sec:p2SEMD}

In this section, we provide a review of the spectral function, define the general SEMD metric from it, and then specialize to the Riemannian $p = 2$ case, for which the SEMD metric can be written in closed form.
Because the explicit form of the $p=2$ SEMD metric is new, we present a detailed discussion of its derivation, but point the reader to \Reference{Larkoski:2023qnv} for more details about the spectral function and its one-dimensional optimal transport.

\subsection{Review of the Spectral EMD}

The spectral function $s(\omega)$ is a one-dimensional representation of a particle collision event or jet.
It is a distribution of pairwise particle angles $\omega$, weighted by the product of energies of the particles in each pair:
\begin{align}
s(\omega)\equiv \sum_{i,j\in {\cal E}} E_i E_j\,\delta(\omega - \omega_{ij}) = \sum_{i\in {\cal E}}E_i^2\delta(\omega) + \sum_{i<j \in {\cal E}}2E_iE_j\,\delta(\omega - \omega_{ij})\,,
\end{align}
where ${\cal E}$ is the set of particles of interest, $\omega_{ij}$ is an appropriately-defined angle or distance between particles $i$ and $j$, and $E_i$ is the energy of particle $i$.
Many plots showing how a spectral function can be constructed from an event or jet will be shown in \Secs{specshapes}{compstudies}.
We also define the cumulative spectral function:
\begin{align}
    S(\omega) = \int_0^\omega d\omega'\, s(\omega'),
\end{align}
and the inverse cumulative spectral function $S^{-1}(E^2)$.
The spectral function is normalized to integrate to the total squared energy:
\begin{align}
E_\text{tot}^2 = S(\omega_{\text{max}}).
\end{align}

The spectral function is IRC safe because it is multi-linear in particle energies and the arguments of the $\delta$-functions are purely functions of angle (with no energy weighting).  
It is also invariant to isometries of the ground space, such as rotations about the colliding beam directions or permutations of particle labels, because it is sensitive only to relative angles. The spectral function almost uniquely defines the positions of particles on the celestial sphere, up to isometries, because both relative angles and particle energies are drawn from continuous distributions on phase space \cite{BOUTIN2004709}.%
\footnote{This is in contrast to distributions on a regular lattice, where pairwise distances are likely to be non-unique.}
By ``almost uniquely'', we mean that there is 0 probability that a spectral function corresponds to two or more distinct, non-isometric distributions of particles on the celestial sphere.
This is necessary to define a metric -- the distance between spectral functions should be always zero if and only if they are the same spectral function, and thus the distance between their corresponding events being zero implies they are \emph{almost} always the same event.
Note that the existence of degenerate configurations is highly nontrivial, and we discuss this point more in \Sec{degeneracies}.

Then, as the fundamental object that encodes complete information about the set of particles, a metric between two events $A,B$ can be defined through the optimal transport of one event's spectral function to the other's.
Further, because the spectral function is one-dimensional, the optimal transport cost can be expressed as a one-dimensional integral of the difference between inverse cumulative spectral functions:
\begin{align}
\text{SEMD}_{p}(s_A, s_B) \equiv \int_0^{E_{\text{tot}}^2} dE^2\, \left|
S_A^{-1}(E^2)-S_B^{-1}(E^2)
\right|^p\,,\label{eq:definition}
\end{align}
for some $p \geq 1$.
Technically, to be a metric (and satisfy the triangle inequality between three events $A,B,$ and $C$), we further need to take the $1/p$ power of the result, but we instead work with this $p$th power to avoid carrying around clunky roots.  $S_A^{-1}(E^2)$ is the inverse cumulative spectral function of event $A$, for which the cumulative spectral function is:
\begin{align}
S(\omega)\equiv \int_0^\omega d\omega'\, s(\omega') = \sum_{i\in {\cal E}}E_i^2 + \sum_{i<j \in {\cal E}}2E_iE_j\,\Theta(\omega - \omega_{ij})\,.
\end{align}

In \Reference{Larkoski:2023qnv}, the $p=1$ metric was studied in detail, which corresponds to the historical Monge-Kantorovich distance or Earth Mover's Distance \cite{monge1781memoire} of the more general class of Wasserstein metrics \cite{kantorovich1939mathematical,wasserstein1969markov,dobrushin1970prescribing}.
In the $p=1$ case only, the inverse cumulative spectral functions can be replaced by the cumulative spectral functions themselves:
\begin{align}
\text{SEMD}_{p=1}(s_A, s_B) \equiv \int_0^{\omega_{\max}} d\omega\, \left|
S_A(\omega)-S_B(\omega)
\right|\,.
\end{align}
For this reason, the $p=1$ metric was previously studied in detail, while $p> 1$ metrics (which require the inverse cumulative spectral function to be defined) were not.
We next demonstrate, however, that this restriction was neither necessary nor especially convenient because we can rather easily exactly evaluate the SEMD integral for $p = 2$.

\subsection{Closed-Form Evaluation}

In this work, we focus exclusively on the $p=2$ SEMD:
\begin{align}\label{eq:semd2start}
\text{SEMD}_{p=2}(s_A, s_B) &= \int_0^{E_\text{tot}^2} dE^2\left|
S_A^{-1}(E^2)-S_B^{-1}(E^2)
\right|^2\\
&=\int_0^{E_\text{tot}^2} dE^2\left(
S_A^{-1}(E^2)^2+S_B^{-1}(E^2)^2-2 S_A^{-1}(E^2) S_B^{-1}(E^2)
\right)
\nonumber\,.
\end{align}
We assume that the two events $A$ and $B$ have the same total energy $E_{\rm tot}$.
(As discussed in \Reference{Larkoski:2023qnv}, this can be accomplished by adding any deficit energy to the less energetic event's spectral function a distance $\omega_R$ away, where $\omega_R$ is a chosen angular scale.)
Here, the inverse cumulative spectral function can be evaluated from:
\begin{align}
S^{-1}(E^2)=\int_0^{\omega_{\max}} d\omega\, \Theta\left(E^2 - S(\omega)\right)\,.
\end{align}
For two events or jets $A,B$ that each contain a discrete number of particles, the SEMD can be evaluated exactly in closed form.
We consider each term in the integrand of \Eq{semd2start} separately, and then add them together appropriately to evaluate the spectral metric.

We begin by evaluating terms whose integrands are squares of the inverse cumulative spectral functions.
To do these integrals, we make a change of variables from squared energy to cumulative spectral function:
\begin{align}
E^2 = S(\omega)\,.
\end{align}
We then find: 
\begin{align}
\int_0^{E_\text{tot}^2} dE^2\, S^{-1}(E^2)^2 &= \int_0^{\omega_{\max}} d\omega\, \omega^2\,s(\omega)\,,
\end{align}
which is just the second moment of the spectral function.  This moment is trivial to evaluate because the angular dependence in the spectral function is in $\delta$-functions:
\begin{align}\label{eq:qsevent}
\int_0^{E_\text{tot}^2} dE^2\, S^{-1}(E^2)^2 &=\sum_{i<j\in{\cal E}}2E_iE_j\omega_{ij}^2\,,
\end{align}
where the sum runs over all particles $i<j$ in the event ${\cal E}$.  This is just a two-point energy correlation function observable \cite{Banfi:2004yd,Larkoski:2013eya} evaluated on the event of interest.

\begin{figure}[t!]
\begin{center}
\includegraphics[width=9cm]{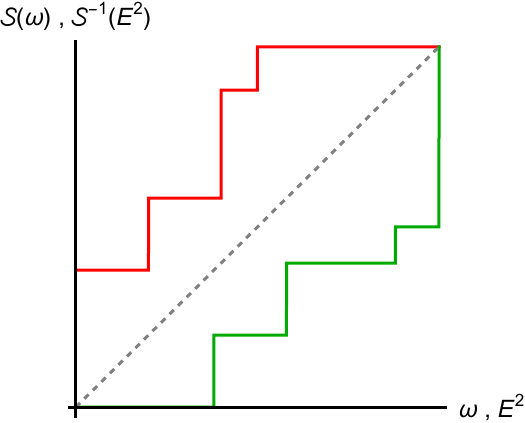}
\caption{\label{fig:invcumspec}
Illustration of the relationship between the cumulative spectral function and its inverse.  The cumulative spectral function $S(\omega)$ (red) takes the value of a squared energy $E^2$ at every angle $\omega$, while the inverse cumulative spectral function $S^{-1}(E^2)$ (green) takes the value of an angle at every squared energy.
}
\end{center}
\end{figure}

Moving on to the mixed term, we can evaluate it from the explicit form of the inverse cumulative spectral function, $S^{-1}(E^2)$.
To do this, it is convenient to introduce some notation.
The cumulative spectral function can be written as:
\begin{align}
S(\omega) = \sum_{i\in{\cal E}} E_i^2+\sum_{i<j\in{\cal E}}2E_iE_j\Theta(\omega-\omega_{ij}) \equiv \sum_{i\in {\cal E}} E_i^2+\sum_{\substack{n\in{\cal E}^2\\\omega_n<\omega_{n+1}}}(2EE)_n\Theta(\omega-\omega_n)\,.
\end{align}
In the sum on the right, $\omega_n$ is one of the ${N\choose 2}$ pairwise angles between the $N$ particles in the event and $(2EE)_n$ is its corresponding squared energy weight.
The notation $n \in{\cal E}^2\equiv {\cal E}\times{\cal E}$ means that $n$ represents a pair of particles from event ${\cal E}$.
In the sum, we have ordered the angular factors so that $\omega_n< \omega_{n+1}$.
We then define \emph{inclusive} and \emph{exclusive} cumulative distributions up to a peak $n\in{\cal E}^2$ as:
\begin{align}
S^+(\omega_n) &\equiv \lim_{\epsilon\to 0} S(\omega_n+\epsilon) = \sum_{i\in {\cal E}} E_i^2+\sum_{\substack{n\geq m\in{\cal E}^2\\\omega_m<\omega_{m+1}}}(2EE)_m\,,\\
S^-(\omega_n) &\equiv \lim_{\epsilon\to 0} S(\omega_n-\epsilon) = \sum_{i\in {\cal E}} E_i^2+\sum_{\substack{n> m\in{\cal E}^2\\\omega_m<\omega_{m+1}}}(2EE)_m\,.
\end{align}
Note that $S^+(\omega_n)$ is the cumulative spectral function including the peak at $\omega = \omega_n$, while $S^-(\omega_n)$ does not include the peak at $\omega_n$. 
Using this notation, the inverse cumulative distribution $S^{-1}(E^2)$ can be expressed as:
\begin{align}
\label{eq:simpler_inverse_cumulative_distribution}
S^{-1}(E^2) = \sum_{\substack{n\in{\cal E}^2\\ \omega_n<\omega_{n+1}}}\omega_n\,\Theta\left(
S^+(\omega_n)-E^2
\right)\Theta\left(
E^2 - S^-(\omega_n)
\right)\,.
\end{align}
The proof of this statement follows simply by sketching the functional form of the cumulative spectral function as in \Fig{invcumspec}, reflecting about the diagonal, and building the resulting function out of $\Theta$-functions.

To express the mixed term, is it further convenient to define the quantity:  
\begin{align}
\label{eq:theta_function}
\mathcal{S}_{n\ell} \equiv \min\left[ S_A^+(\omega_n),S_B^+(\omega_{\ell})
\right]-\max\left[ S_A^-(\omega_n),S_B^-(\omega_{\ell})
\right]\,,
\end{align}
where $A$ and $B$ refer to two \emph{different} spectral functions, indexed by $n$ and $\ell$ respectively.
Integrating the product of spectral functions defined by \Eq{simpler_inverse_cumulative_distribution} and keeping careful track of the $\Theta$-functions, the cross term integral in the SEMD can be written as:
\begin{align}
\int_0^{E_\text{tot}^2}dE^2\, S_A^{-1}(E^2)S_B^{-1}(E^2)= \sum_{\substack{n\in {\cal E}_A^2,\,\ell\in{\cal E}_B^2\\ \omega_n<\omega_{n+1}\\ \omega_{\ell}<\omega_{\ell+1}}}\omega_n \omega_{\ell}\, \text{ReLU}(\mathcal{S}_{n\ell}),
\end{align}
where 
\begin{align}
    \text{ReLU}(x) \equiv x \, \Theta(x)\,,
\end{align}
is the Rectified Linear Unit function~\cite{agarap2019deeplearningusingrectified}.

Summing up the squared and mixed contributions yields a compact form for the SEMD:
\begin{align}
\label{eq:discrete_spectral_emd}
\text{SEMD}_{\beta,p=2}(s_A, s_B) &= \sum_{i<j\in{\cal E}_A}2E_iE_j\omega_{ij}^2+\sum_{i<j\in{\cal E}_B}2E_iE_j\omega_{ij}^2-2\sum_{\substack{n\in {\cal E}_A^2,\,\ell\in{\cal E}_B^2\\ \omega_n<\omega_{n+1}\\ \omega_{\ell}<\omega_{\ell+1}}}\omega_n \omega_{\ell}\,\text{ReLU}(\mathcal{S}_{n\ell})\,. 
\end{align}
We emphasize that this is an \emph{exact} and closed-form expression for the $p = 2$ $\SEMD$ between any two sets of discrete points. 
Moreover, \Eq{discrete_spectral_emd} is extremely fast to evaluate; we will see in \Sec{specter} that it is possible to evaluate it in $\O(N^2\log N)$ time using the highly efficient, differentiable, and parallelizable \Specter framework.
This is in contrast to the $p = 1$ $\SEMD$~\cite{Larkoski:2023qnv}, the ordinary $\EMD$, or approximations to Wasserstein such as sliced-Wasserstein \cite{bonneel2015sliced} or Sinkhorn \cite{cuturi2013sinkhorn}, which either have no closed-form expression, are expensive to evaluate, or are inexact.
Lastly, the structure of \Eq{discrete_spectral_emd} is valid for any choice of ground metric $\omega$. 
In fact, it is not even necessary that $\omega$ be a proper metric at all -- it suffices for $\omega_{ij}$ to be any symmetric matrix such that $\omega_{ii} = 0$.

For events $A$ and $B$ each with $N$ particles, the first two terms in this sum have $\O(N^2)$ terms to evaluate.
The cross term is a sum over both events, and naively appears to have $\O(N^4)$ terms.
However, the structure of $\Theta(\mathcal{S}_{n\ell})$ within the ReLU function ensures that almost all of these terms are zero, and that only $\O(N^2)$ actually contribute to the cross term.
This term takes the form:
\begin{align}
\label{eq:theta_forms}
\Theta(\mathcal{S}_{n\ell})= \Theta\left(
S_A^+(\omega_n) \geq S_B^+(\omega_{\ell}) > S_A^-(\omega_n)
\right)\text{ OR  }\Theta\left(
S_B^+(\omega_{\ell}) \geq S_A^+(\omega_n) > S_B^-(\omega_{\ell})\right).
\end{align}
Here, we have slightly abused the step function notation, with $\Theta(\text{TRUE})  =1$ and $\Theta(\text{FALSE}) = 0$. 
These two inequalities can be  computed efficiently in $\O(N^2 \log N)$ time and makes the SEMD much easier to calculate, as we explored further in \Sec{specter}.

\subsection{Choice of Ground Metric}\label{sec:metrics}

The distance metric $\omega$ depends on the geometry being considered. 
For jets at a hadron collider, we consider a local rectangular patch in the rapidity-azimuth $(y,\phi)$ plane and use the Euclidean metric:\footnote{This differs from the convention of \Reference{Ba:2023hix}, which for $\beta = 2$ has an overall factor of $\frac{1}{2}$ in the metric.} 
\begin{align}
    \label{eq:metric_patch}
    \omega_{ij} = \sqrt{ (y_i - y_j)^2+(\phi_i - \phi_j)^2}.
\end{align}
In this case, we use the particle's transverse momentum $p_T$ in place of its energy $E$, but for the sake of notational simplicity, we continue to use $E$ everywhere in this paper.

For full events in a spherical detector, we use the spherical arc length (angle $\theta$) metric:
\begin{equation}
    \label{eq:metric_sphere}
    \omega_{ij} = \left|\theta_{ij}\right| = \left|\cos^{-1}\left(1-\frac{p_i \cdot p_j}{E_i E_j}\right)\right|,
\end{equation}
where throughout this paper we assume that particle momentum four-vectors $p_i,p_j$ are massless.

There are many other possible geometries and metrics one could have chosen instead, such as cylindrical geometries, chord lengths rather than arc lengths, or even arbitrary warpings of the above metrics, but we stick to these two for the reminder of this paper for simplicity. 
These two metrics in particular are relatively simple to work with and make it easy to relate the SEMD to the EMD and other well-known observables. 
We briefly discuss using the spherical chord length rather than arc length in \App{thrust_chord}.  

We note that the SEMD and \Eq{discrete_spectral_emd} are valid and interpretable even outside the context of collider physics. 
For example, in images, the metric $\omega$ might be a pixel distance, and the energy $E$ might be pixel intensities. 
In general, the SEMD is a valid metric on the space of \emph{arbitrary} weighted distributions -- in collider physics, an event (jet) is a distribution of points $x_i$ with weights $E_i$ ($p_{T,i}$) and pairwise distances $\omega_{ij}$.
These distributions need not even be discrete: the definition of the SEMD given in \Eq{definition} is valid even if $s_A$ or $s_B$ are continuous, though it can no longer be evaluated using the exact expression in \Eq{discrete_spectral_emd} and must be approximated using sampling.

\subsection{Comparison to the EMD and Degeneracies}\label{sec:degeneracies}

The $p =2$ SEMD bears many similarities to the ordinary $\beta = 2$ EMD, and their differences can be illuminating.
The $\beta = 2$ EMD between two events $\E$ and $\E'$, assuming they have the same total energy $E_{\text{tot}}$, is given by~\cite{Komiske:2019fks}:\footnote{For the rest of this paper, whenever we refer to the ordinary EMD, we always mean the $\beta = 2$ EMD.} 
\begin{align}
    \EMD(\E, \E') &= \min_{\pi_{ij} \geq 0}\sum_{ij} \pi_{ij}\,\omega_{ij}^2, \text{ such that } \sum_{i}\pi_{ij} = \frac{E'_j}{E_{\text{tot}}} \text{ and } \sum_j \pi_{ij} = \frac{E_i}{E_{\text{tot}}},
\end{align}
where $i$ indexes particles in $\E$ and $j$ indexes particles in $\E'$.
Both the EMD and SEMD exhibit the same scaling: bilinear in particle energies and quadratic in distance scales, and are expected to be similar in the soft/collinear limits of QCD~\cite{Larkoski:2023qnv}.
In spite of this, their behavior can sometimes differ wildly, especially for ``near-degenerate'' events.

To set the stage, we first review a case where the SEMD and EMD are expected to agree exactly.
Consider a generic jet $\E$ and a special jet $\E'$ consisting of a single particle, such that the center-of-energy $x_0 = \frac{1}{E_{\rm tot}}\sum_i E_i x_i$ of both jets is the same, where $x_i = (y_i, \phi_i)$ is the detector coordinate of particle $i$ and we use the Euclidean metric of \Eq{metric_patch}.\footnote{The center-of-energy is not well-defined for a full event on the celestial sphere, but is well-defined for a jet living on a rapidity-azimuth patch.}
The EMD is straightforward to calculate, since the optimal transport plan is to simply transfer all of $\E$ to the single point of $\E'$:
\begin{align}
    \EMD(\E, \E') &= \frac{1}{2}\frac{\sum_i E_i (x_i - x_0)^2}{\sum_i E_i} 
    = \frac{1}{2}\frac{\sum_{ij} E_i E_j (x_i - x_j)^2}{E_{\rm tot}^2}.
\end{align}
where in the last step, we have used that $x_0 = \frac{1}{E_{\rm}}\sum_j E_j x_j$ is the center-of-energy.
Note that while the $\EMD$ naively appears to be linear in energy, it is actually quadratic once the center-of-energy is accounted for. 
One can show that the $\beta = 2$ EMD between any two localized jets is minimized whenever they share their center-of-energy  -- in this case, the minimization is equivalent to the $\beta = 2$ recoil-free jet angularity~\cite{Berger:2003iw, Berger:2004xf}, the $\beta = 2$ 1-subjettiness~\cite{Thaler:2010tr}, or the yet-to-be-defined 1-sPronginess (see \Sec{1_spronginess} for more details).

We can also calculate the SEMD of this configuration directly using \Eq{discrete_spectral_emd}:
\begin{align}
    \SEMD(s_\E, s_{\E'}) = \sum_{i,j} E_i E_j (x_i - x_j)^2.
\end{align}
Up to the total energy normalization, which we may set to 1, and an unimportant extra factor of 2 (which is due to the $\frac{1}{\beta}$ convention in \Reference{Ba:2023hix}), these agree.
Moreover, both metrics are IRC safe, and therefore continuous on the space of events. 
This implies that for jets $\E'$ that are \emph{almost} one particle, e.g.~one particle plus a soft or collinear emission, $\EMD(\E, \E')$ and $\SEMD(s_{\E}, s_{\E'})$ should be \emph{almost} the same.
Both the $\EMD$ and $\SEMD$ scale bilinearly in energy and scale with $(x_i - x_j)^2$, so one should expect them to be similar.

On the other hand, there are cases where the ordinary $\EMD$ and the spectral $\SEMD$ are very different.
Consider the case where $\E$ is an equilateral triangle of three particles, with one particle having an energy of $\frac{2}{3}$ units, and the other two each having energies of $\frac{1}{6}$ units. 
Without loss of generality, choose the most energetic particle to be at the origin, and one of the two others to be at $(1,0)$. 
Then, choose $\E'$ to consist of two particles, each of energy $\frac{1}{2}$ units, with one at the origin and the other at $(1,0)$. 
Clearly, the energy flows of $\E$ and $\E'$ are different, and so the $\EMD$ will be nonzero. 
However, this specific configuration happens to be a rare degenerate case, with both $\E$ and $\E'$ having the exact same spectral function:
\begin{align}
    s_{\E}(\omega)= s_{\E'}(\omega) = \frac{1}{2}\delta(\omega) + \frac{1}{2}\delta(\omega- 1),
\end{align}
and therefore the $\SEMD$ is zero. 
That is, these two configurations that are very different in ``energy flow space'' become the same event in ``spectral space''.

\begin{figure}[t!]
    \centering
    \subfloat[]{
        \includegraphics[width=0.5\textwidth]{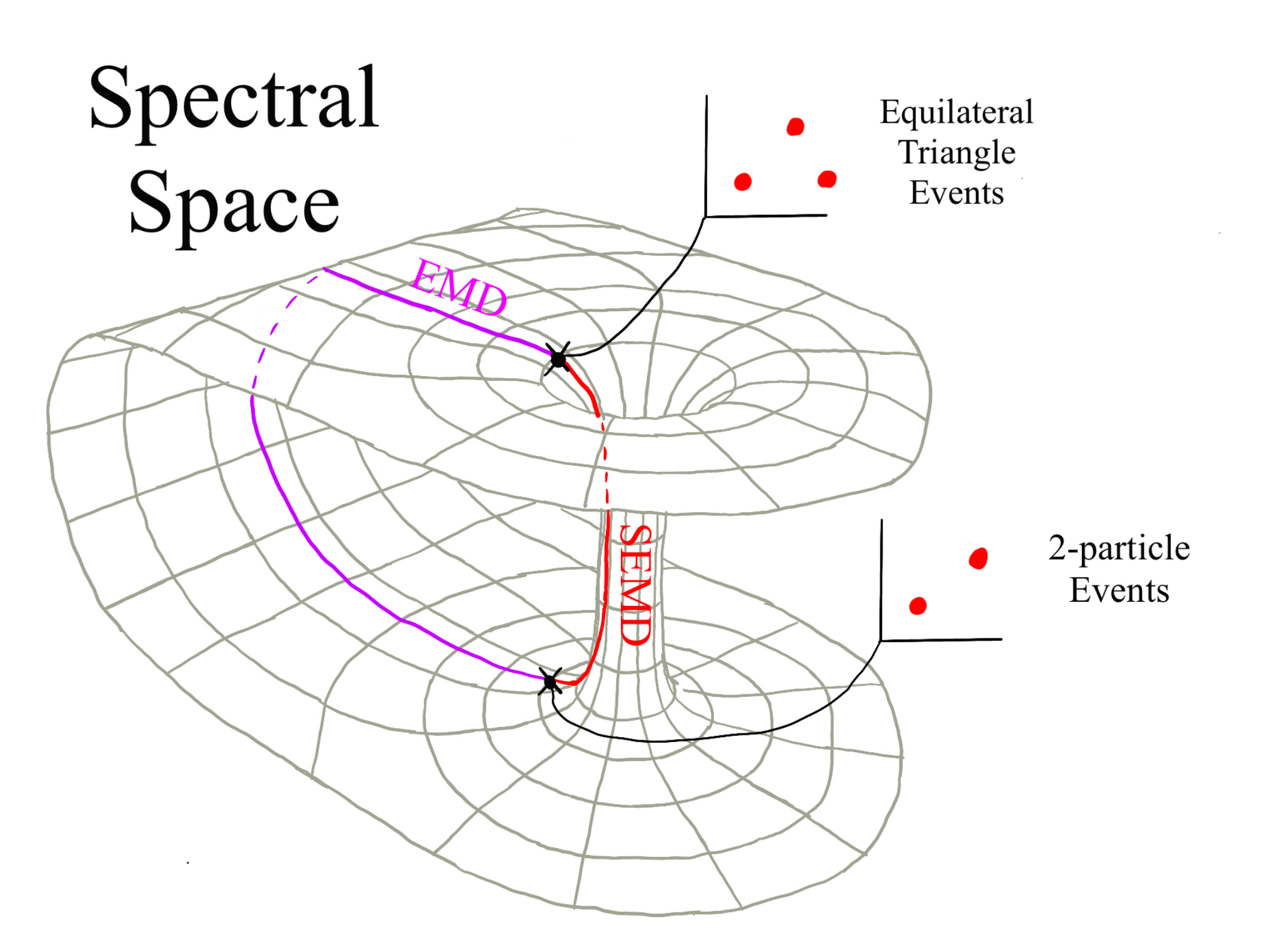}
    }
    \subfloat[]{
        \includegraphics[width=0.45\textwidth]{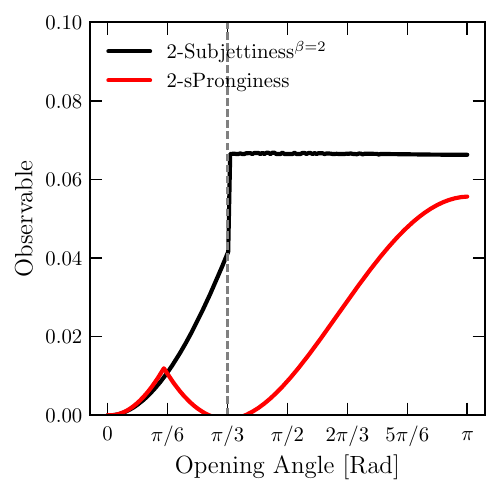}\label{fig:degeneracy_plot}
    }
    \caption{(a) Diagram showing how spectral space induces a different topology on the space of events, as equilateral triangles are closer to 2-particle events in spectral space than in full event space. (b) The minimum EMD (black) and minimum SEMD (red) of an isosceles triangle to all possible 2-particle events (also known as the 2-subjettiness and 2-sPronginess, respectively, defined further in \Sec{n_spronginess}). The isosceles triangle's opening angle is varied. The point of the triangle where the angle varies has an energy of $2/3$ and the other two points each have an energy of $1/6$.}
    \label{fig:degeneracy}
\end{figure}

Even though fully degenerate configurations are just a measure-0 subset of all events, this quirk has implications for the full space of events.
Since both metrics are IRC safe, events that are \emph{almost} degenerate will be separated by a large distance in energy flow space, but will be separated by an \emph{almost} zero distance in spectral space.
The topology induced by the $\SEMD$ is genuinely different than the topology induced by the $\EMD$ on the space of events. 
Observables defined using the $\EMD$ are IRC safe in the sense that the $\EMD$ changes very little under a soft or collinear splitting.  Observables defined using the $\SEMD$ satisfy a slightly stronger condition: the $\SEMD$ changes very little under a soft or collinear splitting \emph{or} a hard splitting that happens to be nearly degenerate. 

In \Fig{degeneracy}, we show the minimum SEMD and EMD between isosceles triangles and the set of all two particle events, also known as the 2-sPronginess (see \Sec{n_spronginess}) and the 2-subjettiness \cite{Thaler:2010tr}, respectively.
The energies of the three particles are as described above, with the apex having an energy of $\frac{2}{3}$, and the opening angle $\theta$ is varied.
The SEMD and EMD agree locally near the $\theta = 0$ configuration, where the event is essentially two particles, and both exhibit the expected quadratic scaling as the opening angle increases.
However, near $\theta = \frac{\pi}{3}$, which is the equilateral triangle configuration, the two sharply diverge, as the event appears to be 3 particles in event space but 2 particles in spectral space.
The agreement between the SEMD and EMD has been ``spoiled'' by the presence of this degeneracy, where otherwise the naive scaling would suggest that the SEMD should continue to grow quadratically alongside the EMD.

Given that the $\EMD$ and $\SEMD$ exhibit similar energy and distance scaling, their differences can be ``blamed'' on these topological differences. 
How often, then, do these differences occur?
The simplest possible degenerate configuration is the equilateral triangle configuration described above.
Generically, any set of three particles whose distances from each other are all $\mathcal{O}(1)$ of each other and whose relative energies are also all $\mathcal{O}(1)$ will be ``almost'' degenerate, and these terms in the spectral function will almost collapse to two points. 
In a light quark or gluon-initiated QCD jet, this is unlikely to occur: emissions of the hard core are strongly ordered in both the emission angle and energy -- that is, QCD jets tend to have 1 hard prong.
Even if a degenerate triangle does occur between three particles of similar energy, their contribution to the spectral function will be highly energy-suppressed relative to the harder component.
In contrast, a top-quark-initiated jet tends to form three hard subprongs in an approximately triangular formation.
Therefore, one should expect that the $\EMD$ and $\SEMD$ should be similar between pairs of QCD jets, but be significantly different between pairs of top jets. 
This is indeed the case; the empirical study in \Sec{empirical_pairwise} shows that the $\EMD$ and $\SEMD$ is indeed similar for QCD jets, but differs on other phase spaces.

\section{The SPECTER Framework}\label{sec:specter}

\Specter is a numeric framework for computing the $p = 2$ $\SEMD$ between events using \Eq{discrete_spectral_emd}, and for computing observables derived from the $\SEMD$ via gradient descent as described further in \Sec{specshapes}.
\Specter is implemented in \textsc{JAX}~\cite{jax2018github}, a Python library for differentiable and compilable computing.
In \Sec{algorithm}, we discuss the algorithm to efficiently compute \Eq{discrete_spectral_emd}.
Then, in \Sec{runtimes}, we demonstrate the speed of the \Specter framework with empirical studies. 
This section is self-contained, and readers less focused on the implementation of the SEMD can safely skip to \Sec{specshapes}.

\subsection{Algorithm for the Spectral EMD}\label{sec:algorithm}

Given an event $\E$ with $N$ massless particles, we first compute its spectral representation $s_\E$. 
We assume that all events are normalized with a total energy $E_{\rm tot} = 1$, and that we can efficiently compute a distance matrix $\omega_{ij}$, computed using either \Eq{metric_patch} or \Eq{metric_sphere}.\footnote{In principle, nothing prevents the use of arbitrary $\omega_{ij}$ in the algorithm, but for simplicity we will stick with the two prescribed metrics.}
The spectral representation is a $2 \times (1 + {N \choose 2})$ array of values, with the first column representing the list of distances $\omega_{ij}$ and the second column representing the list of energies $(2EE)_{ij}$.
The second dimension represents the list of all pairs of particles, plus an additional entry for the $i = j$ pairs. 
We index these pairs by $n$, and we 0-index, with the index $n = 0$ referring to the $i = j$ pairs. 

For $i > j$, the matrix of distances $\omega_{ij}$ are then sorted in ascending order into a single list $\omega_n$ and assigned indices $n = 1, ..., {N \choose 2}$. 
The special value $\omega_0 = 0$ is assigned for $n = 0$.
These values then populate the first column of the $s_\E$ array.
Then, for $n \geq 1$ corresponding to the pair $(i,j)$, the value $(2EE)_{ij} = 2E_iE_j$ is computed and populate the second column, with the special value $(2EE)_0 = \sum_i E_i^2$. 

Importantly, every step of this operation is fully differentiable (even the sorting, as long as the indices are tracked).
The entire operation can be run in $\mathcal{O}(N^2\log N)$ time: the $N^2$ is due to the pairs, and the logarithm to the sorting and searching operations. 
Note that storing the spectral representation of an event requires $\mathcal{O}(N^2)$ in memory, unlike an ordinary event which requires only $\mathcal{O}(N)$ in memory. 
For large $N$, approaching as high as $N \sim 1000$ for heavy ion collision experiments, these memory requirements per event can become a bottleneck of the \Specter algorithm.

We will find it convenient to process many events at once (a \emph{batch}), each with a different number of particles $N$. 
To deal with this and keep all of our arrays rectangular, we choose a fixed $N_{\mathrm{particles}}$.\footnote{In principle, this is not necessary and one can instead use ``awkward arrays''~\cite{Pivarski_Awkward_Array_2018} to avoid the need for rectangular data structures, but for simplicity this is not implemented in v1.0.0 of \Specter.}
Events with fewer particles than this are 0-padded (which will not change the energy flow $\E$ by IRC safety), and events with more particles than this have their least energetic particles removed (which is \emph{not} IRC safe).\footnote{To avoid this, events should be clustered in an IRC-safe way to fewer than $N_{\rm particles}$ ahead of time.}

\begin{figure}[t!]
    \centering
    \includegraphics[width=\textwidth]{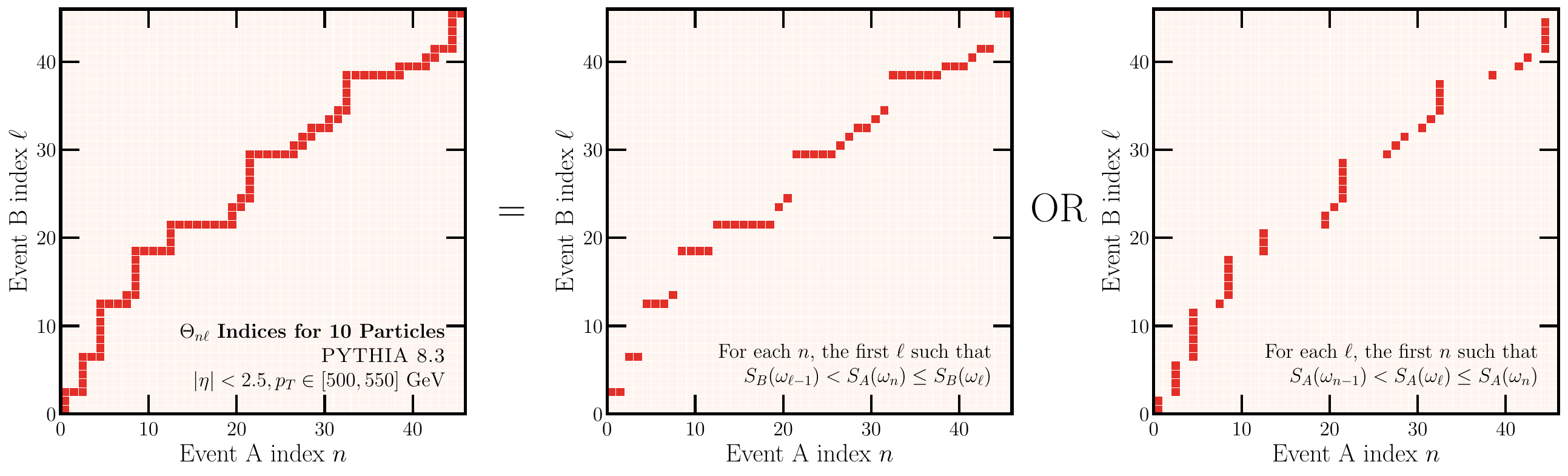}
    \caption{
    (Left) The indices $(n,\ell)$ for which $\Theta(\mathcal{S}_{n\ell})$, as defined in \Eq{theta_function}, is nonzero for a typical pair of events with 10 particles. The bright red points are $(n,\ell)$ pairs where $\Theta(\mathcal{S}_{n\ell}) = 1$. (Middle, Right) This can be computed as the logical OR of \Eq{search_sorted_rels}, which consist solely of searches within sorted lists and are primitive operations within $\textsc{JAX}$.}
    \label{fig:theta}
\end{figure}

Given two events $\E_A$ and $\E_B$, and their spectral representations constructed as described above, we now wish to compute \Eq{discrete_spectral_emd} efficiently.
The first two terms of \Eq{discrete_spectral_emd} are straightforward given the spectral representations $s_\E$, as they take the form:
\begin{align}
    \text{Term 1 + Term 2} &= \int d\omega\, \omega^2 \big(s_A(\omega) + s_B(\omega)\big)\nonumber\\
    &= \sum_{n=0} \omega_n^2 (2EE)_{A,n} + \sum_{\ell=0} \omega_{\ell}^2 (2EE)_{B,n},
\end{align}
which are simple sums with $\mathcal{O}(N^2)$ terms, and they are also straightforward to differentiate through.
The cross term of \Eq{discrete_spectral_emd} is more daunting, as the sum naively has $\mathcal{O}(N^4)$ terms to evaluate.
However, this sum is sparse -- almost all terms are $0$, and only $\mathcal{O}(N^2)$ terms will survive the $\Theta(\mathcal{S}_{n\ell})$ function defined in \Eq{theta_forms}. 
An example of this is shown in \Fig{theta}, which shows $\Theta(\mathcal{S}_{n\ell})$ for a typical pair of events. 
Only a sparse subset of $(n,\ell)$ terms will contribute to the sum.
Moreover, we can efficiently pre-compute which indices $(n, \ell)$ will survive the $\Theta(\mathcal{S}_{n\ell})$ function by taking advantage of the fact that spectral functions are one-dimensional objects that can be sorted.\footnote{This is a concrete realization of the conventional wisdom that having sortable distances is why optimal transport is easier in one dimension than in higher dimensions.}

For a given $(n,\ell)$ pair, \Eq{theta_forms} implies $\Theta(\mathcal{S}_{n\ell})$ is 1 if and only if:
\begin{align}\label{eq:search_sorted_rels}
    S_B(\omega_{\ell-1}) < S_A(\omega_n) \leq S_B(\omega_{\ell}) 
    \qquad{\rm OR} \qquad S_A(\omega_{n-1}) < S_B(\omega_{\ell}) \leq S_A(\omega_n) \,,
\end{align}
where $S(\omega)$ is the cumulative spectral function corresponding to $s(\omega)$.
These relationships are simple searches through sorted lists of the form: for each $n$, the inequality at the left of \Eq{search_sorted_rels} returns an $\ell$ satisfying the inequality, and similarly, for each $\ell$, the inequality at the right of \Eq{search_sorted_rels} returns an $n$ satisfying the inequality. 
This operation is implemented as an efficient and parallelized primitive in many Python libraries -- in $\textsc{JAX}$, it is the $\texttt{searchsorted}$ function, which returns the \emph{first} $\ell$ or $n$ satisfying the inequalities. 
We need to combine the inequalities with a logical OR to find all such $\ell$ or $n$, which is demonstrated in \Fig{theta}. For each $n$, the search operation runs in only $\mathcal{O}(\log N)$ since $S(\omega_{\ell})$ is sorted in $\ell$ by construction, and so for the entire event it only takes $\mathcal{O}(N^2 \log N)$ to find all indices that contribute to the cross term.
The logical OR structure implies that there can be \emph{at most} $2 {N \choose 2} < N^2$  pairs of indices, which allows us to set an upper bound to the allocated memory, though in practice this limit is rarely saturated.
Note that the $n = 0$ and $\ell = 0$ terms can be skipped, since $\omega_0 = 0$ and therefore does not contribute to the sum.

With the list of indices $(n,\ell)$ for which $\Theta(\mathcal{S}_{n\ell})$ is nonzero in hand, the cross term in \Eq{discrete_spectral_emd} is now faster to evaluate, as we only need to sum $\mathcal{O}(N^2)$ terms. 
The indices do not need to be recomputed when calculating derivatives, so time is in fact saved if the same $\Theta$ function is used for both the SEMD and its derivatives with respect to $s_\E$.
Thus, the entire SEMD computation can be performed exactly in $\mathcal{O}(N^2 \log N)$ time and $\mathcal{O}(N^2)$ memory, in a way that is easily amenable to parallelization and differentiation.

It is important to emphasize that this is an \emph{exact} algorithm to compute the $\SEMD$.
No approximations have been made, numerical or otherwise. 
This is important to emphasize since when evaluating spectral shape observables (see \Sec{specshapes} for more details), there are two approximations made: first, continuous events are approximated as discrete, so that this algorithm may be applied; second, evaluating shape observables requires gradient descent over the shape parameters, which is an inexact minimization procedure. 
However, the $\SEMD$ evaluation itself is exact -- this is in contrast to \Shaper, where on top of these two approximations, the $\EMD$ computation is also an approximation using the debiased Sinkhorn divergence \cite{sinkhorn1966relationship,cuturi2013sinkhorn,clason2021entropic,feydy2019interpolating,janati2020debiased}.

\subsection{Runtime Benchmarks}\label{sec:runtimes}

\begin{figure}[t!]
    \centering
    \includegraphics[width=0.8\textwidth]{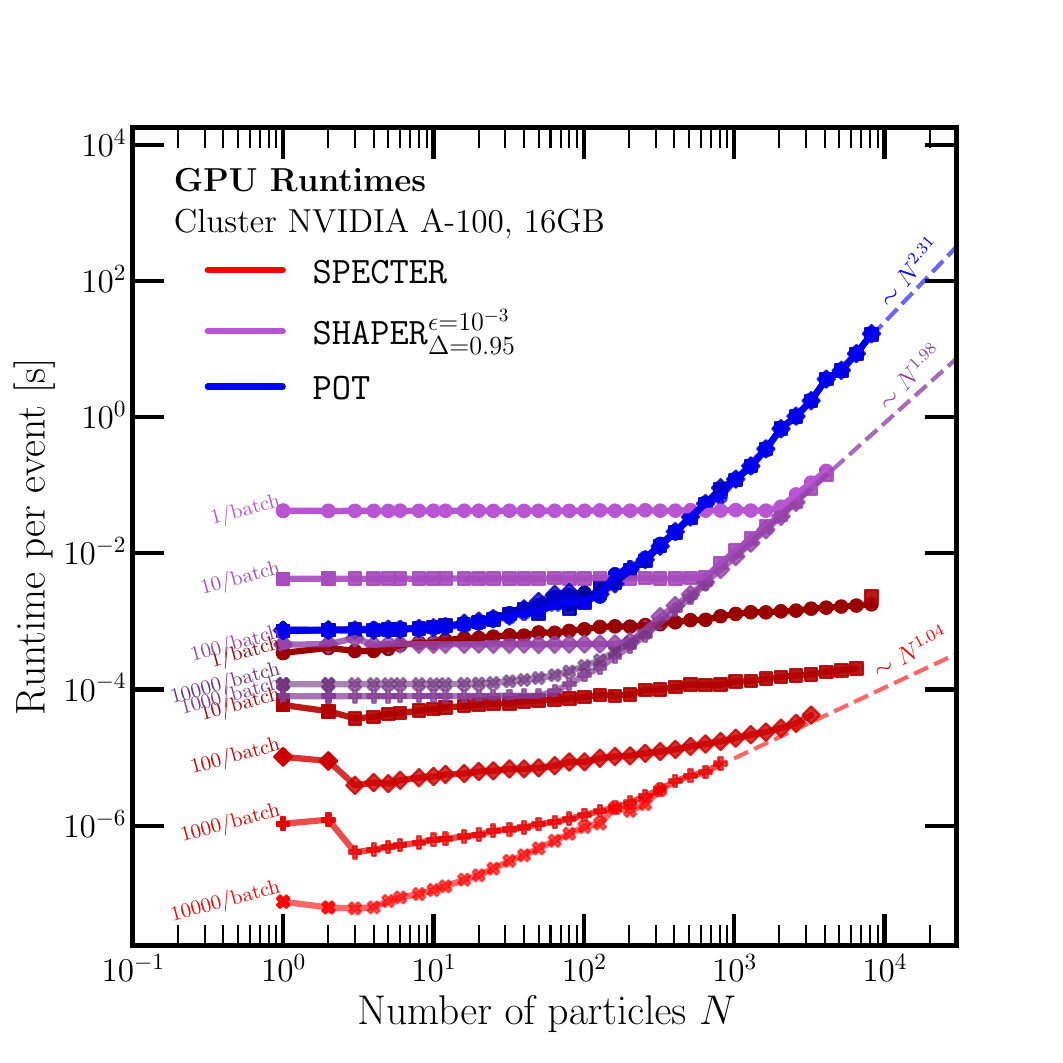}
    \caption{Runtimes for various (S)EMD algorithms as a function of the number of particles $N$ in each event, as run on a NVIDIA A100 GPU with 16 GB of memory. The SEMD, as evaluated by \Specter, is shown in red. The ordinary EMD, as estimated by the Sinkhorn algorithm with \Shaper and in full with the Python Optimal Transport library, are shown in purple and blue, respectively. All lines eventually scale as  $N^{\sim 2}$ or better.}
    \label{fig:GPU_runtimes}
\end{figure}

\begin{figure}[t!]
    \begin{center}
    \includegraphics[width=0.8\textwidth]{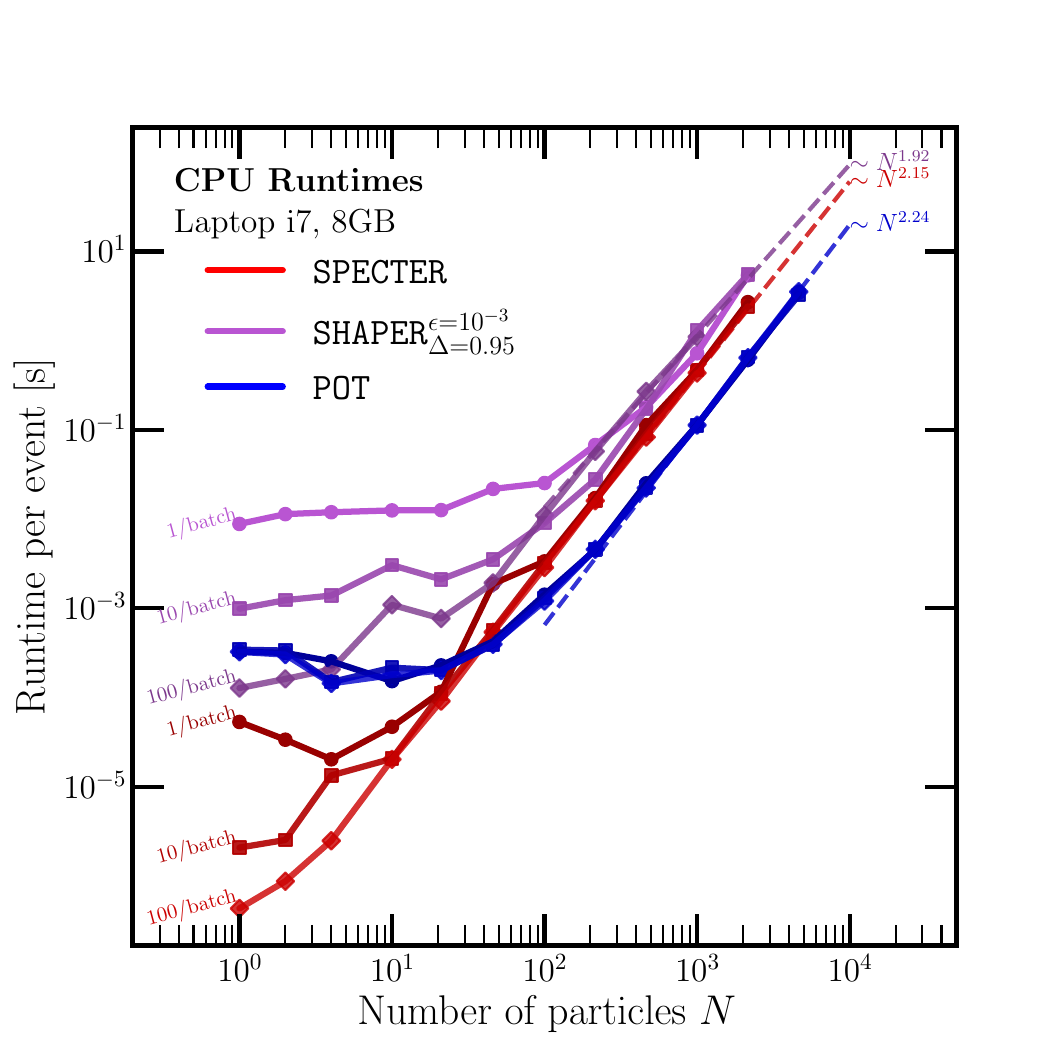}
\caption{\label{fig:CPU_runtimes} The same as \Fig{GPU_runtimes}, but as run on an ordinary laptop CPU with 8 GB of memory.  Each of the three lines corresponds to batch sizes of $1$, $10$, and $100$ respectively. All lines eventually scale as $N^{\sim 2}$.}
    \end{center}
\end{figure}

In \Figs{GPU_runtimes}{CPU_runtimes}, we show the runtime of \Specter to calculate the SEMD between two events with $N$ particles. 
For comparison, we also show the runtime for calculating the ordinary EMD using the approximate Sinkhorn algorithm via \Shaper, and an exact calculation using the Python Optimal Transport library (POT) \cite{flamary2021pot}. 
\Fig{GPU_runtimes} shows the runtimes when running on a compute cluster GPU (specifically, a NVIDIA A100 with 16 GB of memory) and \Fig{CPU_runtimes} shows the runtimes when running on an ordinary laptop CPU (specifically, one of the authors' Dell Inspiron 16 with an Intel i7-1255Ux12 CPU with 8 GB of memory), which together span the typical use cases. 
We include curves for several different batch sizes to show the impact of parallelization: 1, 10, 100, 1000, and 10000 pairs of events at once in \Fig{GPU_runtimes}, and 1, 10, and 100 pairs of events at once in \Fig{CPU_runtimes}.
Note that some of the curves for \Shaper and \Specter truncate early: this is due to memory limitations. 
In particular, because JAX arrays are indexed using int32, and \Specter requires $\O(N^2)$ memory, events with more than $2^{\sim 16} \approx 30000$ particles cannot be handled.

On the GPU, \Specter is \emph{significantly} faster than the ordinary EMD calculation, achieving speeds that are hundreds to thousands of times faster than \Shaper or POT. 
For events with $\sim 100$ particles, which is typical in jet physics, it takes only 0.001 seconds to compute the SEMD between 10000 pairs of events.
The corresponding calculation of the EMD on the same hardware using \Shaper would take $\sim$ 1 second (and it would only be an estimate), and nearly 100 seconds using POT.
Even on a CPU, \Specter is faster than either \Shaper or POT at low $N$, and is approximately just as fast at moderate and large $N$.

All algorithms tend towards $\sim \O(N^2)$ at large $N$.
However, especially on the GPU (\Fig{GPU_runtimes}), the runtime at $N \lesssim 100$ is dominated by an overhead component, and in this regime the scaling is significantly better than $\O(N^2)$.
In fact, an $\O(N)$ or smaller overhead seems to completely dominate the \Specter runtimes across the entire range before being cutoff due to memory limitations, which becomes more apparent with larger batch sizes.
It is possible that this is not saturated in our trials, and that even faster per-event runtimes might be achieved with larger batch sizes.  
In all cases, the compile time is not considered: this is the long-run per-event runtime.
Both \Shaper and \Specter take less than one minute to compile on both the laptop and GPU hardware.


\section{Spectral Shapes}\label{sec:specshapes}

Having defined the $p = 2$ SEMD and presented the \Specter algorithm for efficient evaluation, we now show how the SEMD can be used to define novel \emph{spectral} event and jet shape observables.
These are analogous to observables defined using the ordinary EMD~\cite{Komiske:2019fks, Ba:2023hix}, but with the added benefit of being much easier to evaluate on events due to the speed of \Specter.
In some cases, it is even possible to define observables with tractable closed-form expressions, in contrast to the EMD where it is difficult to find closed-form expressions in all but the most trivial cases.
This makes it possible to define shape observables that both probe event geometry \emph{and} can be evaluated extremely quickly, and in some cases practically instantaneously with an exact closed-form expression.

\subsection{General Aspects}

Following the philosophy of \Reference{Ba:2023hix}, \emph{any} metric on the space of events can be used to define an infinite family of event and jet shape observables.
The SEMD is no different in this regard.
A ``shape'' can be defined as a parameterized energy flow $\E_{\rm Shape}^{(\theta)}$ on detector space, which then induces a corresponding spectral function $s_{\rm Shape}^{(\theta)}$.
Then, for an event or jet with energy flow $\E$ and corresponding spectral function $s_\E$, we can define a \emph{spectral shape observable} $\mathcal{O}_{\rm Shape}(s_\E)$ and the corresponding \emph{spectral shape parameters} $\theta_{\rm Shape}(s_\E)$:\footnote{Note on naming conventions: we use the argument of the function to distinguish the EMD-based observables and parameters from the SEMD-based ones with $\O_{\rm Shape}(\E)$ and $\theta_{\rm Shape}(\E)$ for the ordinary EMD, and $\O_{\rm Shape}(s_\E)$ and $\theta_{\rm Shape}(s_\E)$ for the SEMD. When naming shapes, we use ``s'' or ``sp'' in front to signal the spectral version.}
\begin{align}
    \O_{\rm Shape}(s_\E) &= \min_{\E^{(\theta)}_{\rm Shape}} \SEMD (s_\E, s^{(\theta)}_{\rm Shape}) \label{eq:observable_definition}\\
    \theta_{\rm Shape}(s_\E) &= \argmin_{\E^{(\theta)}_{\rm Shape}} \SEMD (s_\E, s^{(\theta)}_{\rm Shape}).\label{eq:parameters_definition}
\end{align}
Given the algorithm in \Sec{algorithm} for efficiently computing the $\SEMD$ in a differentiable way, one can numerically approximate the (arg)minimum via gradient descent.
This is done essentially identically to \Shaper~\cite{Ba:2023hix}.

Note that even though we use the SEMD, as defined on spectral densities, we still perform the optimization over detector space events $\E^{(\theta)}_{\rm Shape}$ rather than directly over parameterized spectral functions $s^{(\theta)}_{\rm Shape}$.
This is to ensure that the final parameters correspond to a physically realizable event -- not all spectral densities correspond to physical values of energy or particle positions.
For instance, the spectral density:
\begin{align}
    s(\omega) = E_{\tot}^2 \delta(\omega - \omega_0),
\end{align}
for nonzero $\omega_0$ can only come from a non-physical event with two particles, one with energy $E_1 = \frac{E_{\tot}}{2}(1+i)$ and the other with energy $E_2 = \frac{E_{\tot}}{2}(1-i)$, and therefore one cannot directly optimize over $\omega_0$ if one wants to avoid these ``ghost events''.
However, this is still technically a valid optimization to perform, and the observables $\O$ and $\theta$ obtained this way are perfectly well-defined if one is willing to accept the minimization over non-physical ghost events, though we do not do so here.

Given this setup, all one must do to define a shape is to specify a manifold $\mathcal{M}$ of parameterized energy flows $\E^{(\theta)}_{\rm Shape}$ with coordinates $\theta$. 
This is exactly identical to how observables are defined using \Shaper as described in Sec.~3 of \Reference{Ba:2023hix}, and indeed all the same manifolds of energy flows can be carried over. 
Shapes are defined by first specifying the parameterization manifold $\mathcal{M}_{\rm Shape} \ni \theta$, and then providing a sampling procedure to sample $N$ weighted points based on those parameters.
For example, one can define the ``spectral Ringiness'' (or ``spRinginess'') of an event, which is parameterized by a positive radius ($R \in \mathcal{M}_{\rm Ring} = \mathbb{R}_+ $), plus a function to sample $N$ points uniformly on a circle of radius $R$ each with weight $1/N$.
As with \Shaper, multiple primitive shapes can be composed together to form composite shapes that probe more complex geometries.

The isometry-invariance of the SEMD leads to a small subtlety when parameterizing shapes.
Because the SEMD is invariant under translation and rotation of either event, parameters that affect the overall translational or rotational degrees of freedom will carry no information. 
For example, while the ordinary Ringiness observable needs an overall translational degree of freedom to specify the location of the circle, the spRinginess does not.
The same observable definition and sampling can be used for both, but the spectral observable will be completely agnostic to the unneeded parameter.
The rest of this section is dedicated to discussing examples of spectral shape observables.

\subsection{Spectral Prong Shapes}\label{sec:prong_shapes}

The most basic possible spectral shape observables we can define are the $N$-sPronginess observables, which are the spectral equivalent of the $\beta = 2$ $N$-(sub)jettiness~\cite{Thaler:2010tr, Stewart:2010tn, Stewart:2015waa} observables.\footnote{We call it the $N$-sPronginess because ``$N$-ssubjettiness'' [sic] would have been a terrible name.}
These observables probe how much an event looks like $N$ prongs.
These observables can be defined either on local patches of the rapidity-azimuth plane for jets using the metric in \Eq{metric_patch} (corresponding to $N$-subjettiness), or over the entire detector for full events using the metric in \Eq{metric_sphere} (corresponding to $N$-jettiness with no beam parameter).\footnote{It is possible to define a beam parameter by using unbalanced spectral functions, by introducing a scale parameter $\omega_R$ to re-balance the spectral functions as described in Eq. (2.13) of \Reference{Larkoski:2023qnv}.} 

\subsubsection{1-sPronginess}\label{sec:1_spronginess}

The simplest spectral shape observable is the \emph{1-sPronginess}, which probes how much the event or jet looks like it has 1 prong. 
This is accomplished by calculating the (spectral) distance to an event with just a single particle, given by the energy flow density:
\begin{align}
    \E_\text{1-Prong}(x) = E _\text{tot}\delta(x - x_0),
\end{align}
where $E _\text{tot}$ is the total energy of the particle and $x_0$ is its position in the detector.
However, this position does not matter due to the translational invariance of the metric, for which the relevant object is the spectral representation:
\begin{align}
s_\text{1-Prong}(\omega) = E_\text{tot}^2\delta(\omega).
\end{align}
Note that there are no free parameters in the spectral function -- in spectral space, there is only \emph{one} single particle event up to energy normalization.
Given this, we can exactly compute the 1-sPronginess observable for an event $\O_\text{1-Prong}(s_\E)$, as the $\SEMD$ between $s_\E$ and $s_\text{1-Prong}$ by using \Eq{discrete_spectral_emd}: 
\begin{align}
\label{eq:1_spronginess}
\O_\text{1-Prong}(s_\E) = \sum_{i<j\in {\cal E}}2E_iE_j\omega_{ij}^2.
\end{align}
In this form, the 1-sPronginess is actually identical to many other well-known observables.
For example, the 1-sProginess is identical to the recoil-free $\beta = 2$ angularity~\cite{Berger:2003iw, Berger:2004xf}, which is itself identical to the $\beta = 2$ 1-subjettiness~\cite{Thaler:2010tr} for $\omega = \Delta R$ as in \Eq{metric_patch}.

\subsubsection{$N$-sPronginess}\label{sec:n_spronginess}

\begin{figure}[t!]
    \centering
    \subfloat[]{
        \includegraphics[width=0.5\textwidth]{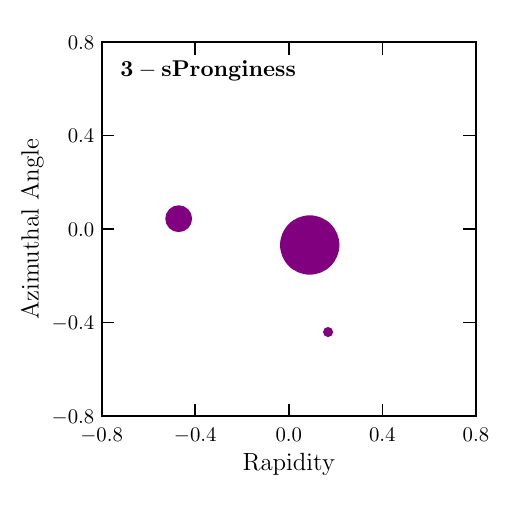}
    }
    \subfloat[]{
        \includegraphics[width=0.485\textwidth]{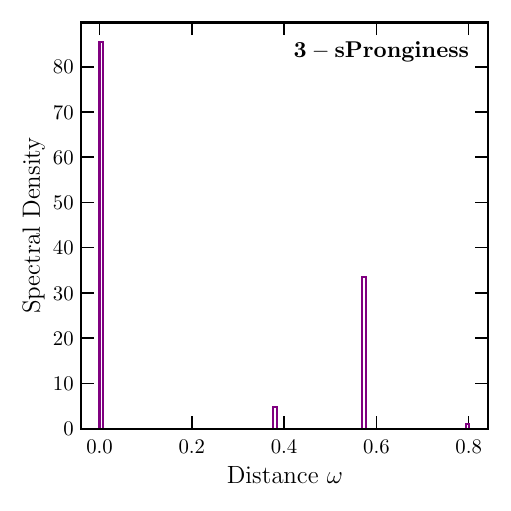}
    }
    \caption{The (a) detector representation from \Eq{n_spronginess_event} and (b) spectral representation from \Eq{n_spronginess_spectral} of an idealized $3$-(s)Prong event, in a patch of the rapidity-azimuth plane. Each dot is a particle, with its size representing its energy weight.}
    \label{fig:example_3sprong}
\end{figure}

The first \emph{nontrivial} spectral observable is the \emph{$N$-sPronginess}, for $N > 1$, which measures how much an event $\E$ looks like it has $N$ prongs. 
These are characterized by calculating the spectral distance to events with $N$ particles:
\begin{align}
    \E_\text{$N$-Prong}^{(E_i, x_i)}(x) = \sum_i^N E_i\, \delta(x - x_i),\label{eq:n_spronginess_event}
\end{align}
where $E_i$ are the particle energies, constrained to be positive and sum to $E_{\rm tot}$, and $x_i$ are the particle positions. 
The corresponding spectral density is:
\begin{align}
s_\text{$N$-Prong}^{(E_i, x_i)}(\omega) = \sum_{i\in {\cal E}_\text{$N$-Prong}} E_i^2\delta(\omega) + \sum_{i<j\in {\cal E}_\text{$N$-Prong}} 2E_iE_j\delta(\omega-\omega_{ij}), \label{eq:n_spronginess_spectral}
\end{align}
where $\omega_{ij}$ is the distance between $x_i$ and $x_j$, as defined using either \Eq{metric_patch} or \Eq{metric_sphere}.
Note that due to translational invariance, one of the $x_i$ degrees of freedom is redundant.
Additionally, there is an overall redundant rotational degree of freedom between the $x_i$'s.
An example of a $3$-(s)Prong event and its spectral representation on the rapidity-azimuth plane is shown in \Fig{example_3sprong}.

Given this, we can define the $N$-sPronginess observable $\O_\text{$N$-Prong}(s_\E)$ and corresponding spectral parameters $\theta_\text{$N$-Prong}(s_\E) = (E_i, x_i)$ via \Eqs{observable_definition}{parameters_definition},\footnote{Up to the redundant translational and rotational degrees of freedom.} which can be numerically evaluated using the \Specter algorithm.
For a given choice of $(E_i, x_i)$ parameters, the $\SEMD$ can be exactly evaluated using \Eq{discrete_spectral_emd}, though the minimization over the $(E_i, x_i)$ is nontrivial.
To choose initial values of $E_i$ and $x_i$ in an IRC-safe way, we simply apply $k_T$-clustering \cite{Catani:1993hr,Ellis:1993tq} to obtain $N$ subjets.
The positivity and sum rule constraints on $E_i$ can be enforced during gradient descent using the simplex projection algorithm~\cite{tankala2023kdeepsimplexdeepmanifold, mueller2022geometricsparsecodingwasserstein, wang2013projectionprobabilitysimplexefficient}.

As discussed in \Sec{degeneracies}, the $N$-sPronginess for $N > 1$ is nontrivial in that it is distinctly different than its corresponding ordinary EMD observable, the $N$-subjettiness, which is defined using the same parameterized energy flows. 
The $N$-sPronginess probes slightly different physics than the $N$-subjettiness: while a low value of $2$-subjettiness implies a jet looks like it has two hard subprongs, a low value of $2$-sPronginess implies that a jet looks like it has two hard subprongs \emph{or} has three hard subprongs in an equilateral triangle formation.

\subsection{Spectral Jet Shapes}\label{sec:jet_shapes}

Building off of the $N$-sPronginess, we can build more nontrivial observables that probe the structure of localized jets. 
In this subsection, we define several novel spectral jet shape observables and discuss their properties.
For all of the observables in this subsection, we take the ground metric $\omega$ to be the Euclidean metric in \Eq{metric_patch}.

\subsubsection{Jet spRinginess}\label{sec:jetringiness}

We define the jet \emph{spRinginess} observable, in analogy with the Ringiness observables of \Reference{Ba:2023hix}.
A ring-shaped event with radius $R$ is given by the energy flow density:
\begin{align}
    \E_{\text{Jet Ring}}^{(R)}(x) = \frac{E_{\tot}}{2\pi R}\, \delta(|x| - R). \label{eq:ring_density}
\end{align}
The radius $R$ is an explicit parameter that will later have to be optimized.
The corresponding spectral representation is:
\begin{align}
    s_{\text{Jet Ring}}^{(R)}(\omega) = \frac{E_\tot^2 }{\pi}\frac{2}{\sqrt{4R^2 - \omega^2}}. \label{eq:ring_spectral}
\end{align}

\begin{figure}[tpb!]
    \centering
    \subfloat[]{
        \includegraphics[width=0.5\textwidth]{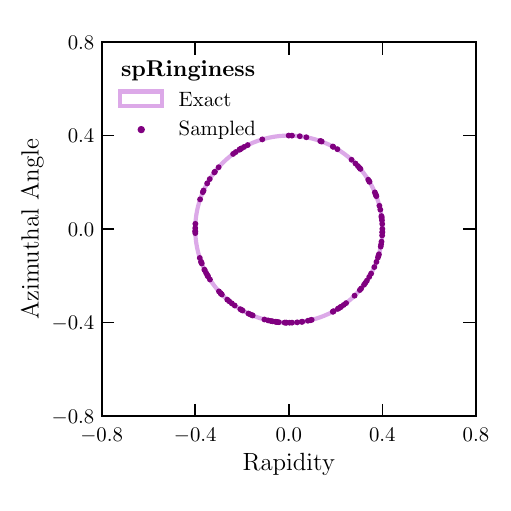}
    }
    \subfloat[]{
        \includegraphics[width=0.485\textwidth]{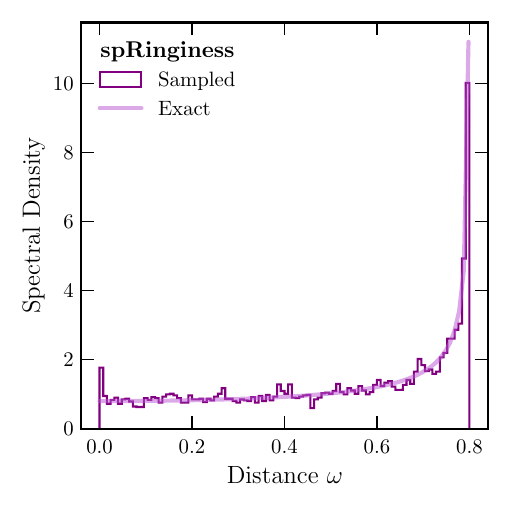}\label{fig:example_ring_sample}
    }
    \caption{The (a) detector representation from \Eq{ring_density} and (b) spectral representation from \Eq{ring_spectral} of an idealized ring jet, in a patch of the rapidity-azimuth plane. Each dot is a sampled particle, with its size representing its energy weight, which happens to be uniform.} 
    \label{fig:example_ring}
\end{figure}
Unlike the $N$-sPronginess of the previous subsection, this is a continuous rather than discrete distribution. However, one can approximate the continuous distribution by simply sampling a large number $N$ of points in a circle of radius $R$, each with energy $\frac{E_{\rm tot}}{N}$. Then, all one has to do is use the \Specter algorithm to compute the $\SEMD$ between the event of interest $\E$ and the discrete event approximating the sphere, and use gradient descent to find the optimal $R$.
There are a variety of ways to choose the initial value of $R$, such as the distance between the leading two subjets, but for simplicity we choose to initialize $R = 0$ so that the observable can be treated as a perturbation to the 1-sPronginess, similar to the choices of initialization for \Shaper. 
Unlike the ordinary jet Ringiness, however, there is little danger of getting stuck in a local minimum, since we will see extremely shortly in \Eq{semd_jet_ring} that the $\SEMD$ is convex in $R$.
An example ring is shown in \Fig{example_ring}.
In \Fig{example_ring_sample}, there is a peak in the sampled ring distribution at $\omega = 0$. 
The $\omega = 0$ bin receives contributions from the individual particle energies squared -- in the fully continuous distribution, these are infinitesimal quantities, but in a uniform sample with $N$ particles, this peak will go as $N\times \frac{1}{N^2}$.
This peak is a generic feature of sampling continuous shapes.

 The jet spRinginess is our first example of a nontrivial observable with a fully closed-form expression: by using the cumulative spectral function and \Eq{semd2start}, we can actually completely bypass the \Specter algorithm and sampling just described and exactly evaluate the jet spRinginess and $R_{\rm opt}$ (see \App{closed_form_jet_springiness} for details). 
The optimal radius $R_{\text{opt}}(s_\E)$ of an event $\E$ with spectral representation $s_\E$ is given by:
\begin{align}
R_\text{opt} = \frac{2}{\pi}\sum_{\substack{n\in{\cal E}^2\\\omega_n<\omega_{n+1}}}\omega_n\left[
\cos\left(
\frac{\pi}{2E_\text{tot}^2}S^-(\omega_n)
\right)-\cos\left(
\frac{\pi}{2E_\text{tot}^2}
S^+(\omega_n)
\right)
\right]\,.
\end{align}
The spRinginess, $\O_{\text{Jet Ring}}(s_\E)$, of the event is then given by:
\begin{equation}
\label{eq:semd_jet_ring}
\O_{\text{Jet Ring}}(s_\E) = \min_R \text{SEMD}_{\beta,p=2}\left(s_\E, s_\text{Jet Ring}^{(R)}\right) = \sum_{i<j\in{\cal E}}2E_iE_j\omega_{ij}^2 - 2E_\text{tot}^2 R_\text{opt}^2\,.
\end{equation}

\subsubsection{Jet spLineliness}\label{sec:jetlineliness}

\begin{figure}[t]
    \centering
    \subfloat[]{
        \includegraphics[width=0.5\textwidth]{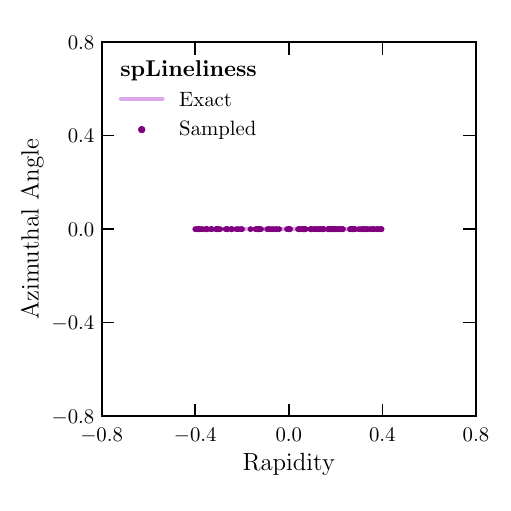}
    }
    \subfloat[]{
        \includegraphics[width=0.485\textwidth]{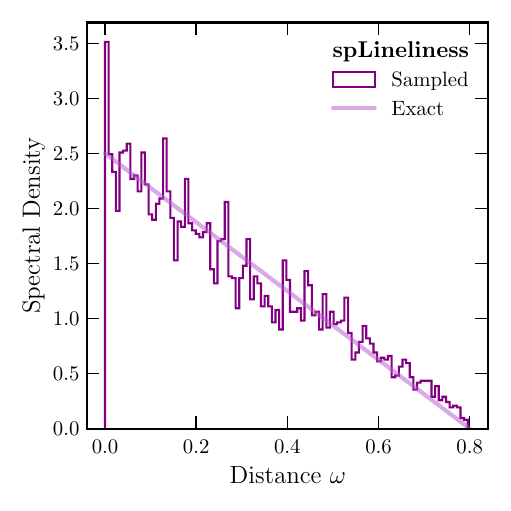}
    }
    \caption{The (a) detector representation from \Eq{line_density} and (b) spectral representation from \Eq{line_spectral} of an idealized line jet, in a patch of the rapidity-azimuth plane.
    Each dot is a sampled particle, with its size representing its energy weight, which happens to be uniform.}
\label{fig:example_line}
\end{figure}
Similar to jet spRinginess, which probes how ring-like a jet is, we can also ask how line-like a jet is to define the jet \emph{spLineliness}.
Here, we compare to line-segment-like events with a length $L$:
\begin{align}
    \E^{(L)}_{\text{Jet Line}}(x)  = \frac{E_{\rm tot}}{L}\, \delta(x^{(2)} - 0)\, \Theta(0 < x^{(1)} < L)\,,  \label{eq:line_density}
\end{align}
where $x^{(1)}$ and $x^{(2)}$ are the first and second detector coordinates, respectively (e.g., in rapidity-azimuth space, $x^{(1)}$ is the rapidity and $x^{(2)}$ is the azimuthal angle).
Up to isometries, this describes all possible line-segment-like jets.
The corresponding spectral representation is:
\begin{align}
     s^{(L)}_{\text{Jet Line}}(\omega)  = \left(\frac{2}{L} - \frac{2\omega}{L^2}\right)E_{\rm tot}^2.  \label{eq:line_spectral}
\end{align}
An example line is shown in \Fig{example_line}.

\begin{figure}[p]
\centering
\subfloat[]{
\includegraphics[width=0.45\textwidth]{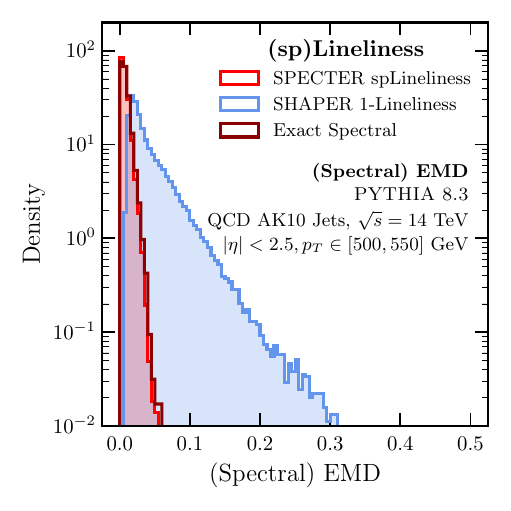}
}\label{fig:line_emd}
\subfloat[]{
\includegraphics[width=0.45\textwidth]{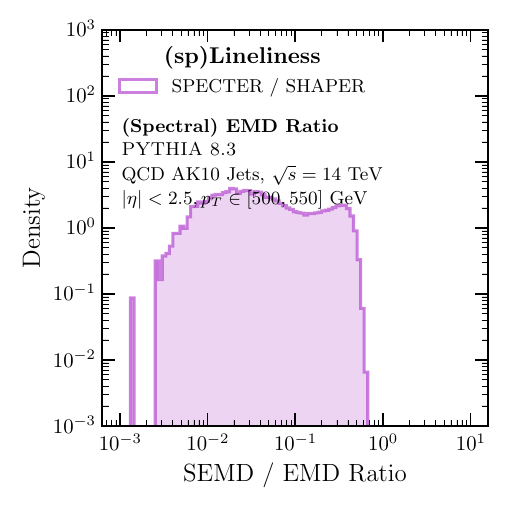}
}\label{fig:line_ratio}
\vspace{1em}
\subfloat[]{
\includegraphics[width=0.45\textwidth]{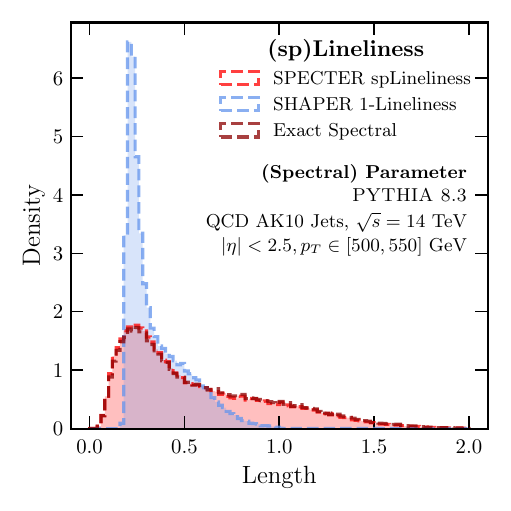}
}\label{fig:line_param}
\subfloat[]{
\includegraphics[width=0.45\textwidth]{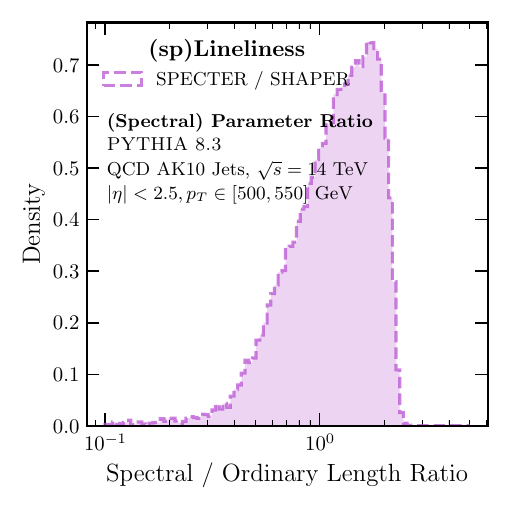}
}\label{fig:line_param_ratio}
\caption{Distributions for the (sp)Lineliness observable, as defined in \Sec{jetlineliness}, evaluated on jets using \Specter (red) and \Shaper (purple). The observable value is shown in (a), and the corresponding shape parameter is shown in (c). The ratio of the spLineliness to the Lineliness is shown in (b) and (d).}
\label{fig:lines}
\end{figure}

Just like the jet spRinginess, both the jet spLineliness and the optimal parameter $L_{\rm opt}$ can be numerically estimated with the \Specter algorithm, by sampling points on a line segment of length $L$ and performing gradient descent over $L$ to minimize the $\SEMD$.%
\footnote{\Specter is built off of and heavily borrows from \Shaper, especially in its sampling procedure for shapes. The original version of \Shaper did not have line segments built in, but it is possible to cheat and sample line segments by using \Shaper's built-in ellipses with $R_1 = L/2$ and $R_2$ frozen to zero. The fact that the exact and numerical curves in \Fig{lines} match shows that this is fine to do, and shows how shapes can be manipulated in both programs.}
Similarly, it is possible to evaluate the spLineliness and optimal length in closed form to bypass the numeric estimate (see \App{closed_form_splineliness} for details):
\begin{align}
L_\text{opt}(s_\E) &= \frac{6}{E_\text{tot}^2}\sum_{\substack{n\in{\cal E}^2\\\omega_n<\omega_{n+1}}}\omega_n\left[
S^+(\omega_n)+\frac{2}{3E_\text{tot}}\left(
E_\text{tot}^2-S^+(\omega_n)
\right)^{3/2}\right.\\
&
\hspace{4.5cm}\left.-\,S^-(\omega_n)-\frac{2}{3E_\text{tot}}\left(
E_\text{tot}^2-S^-(\omega_n)
\right)^{3/2}
\right]\nonumber\,. \\
\label{eq:jetlineobs}
\O_{\text{Jet Line}}(s_\E) &= \sum_{i<j\in{\cal E}}2E_iE_j\omega_{ij}^2-\frac{L_\text{opt}^2}{6}\, E_\text{tot}^2\,.
\end{align}

\subsubsection{Jet sDiskiness}\label{sec:jetdiskiness}

\begin{figure}[t!]
    \centering
    \subfloat[]{
        \includegraphics[width=0.5\textwidth]{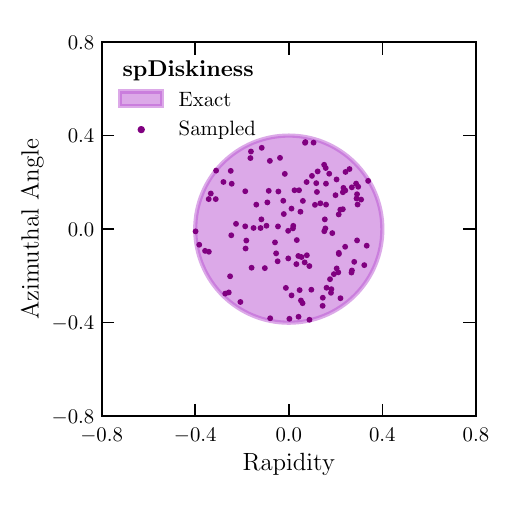}
    }
    \subfloat[]{
        \includegraphics[width=0.485\textwidth]{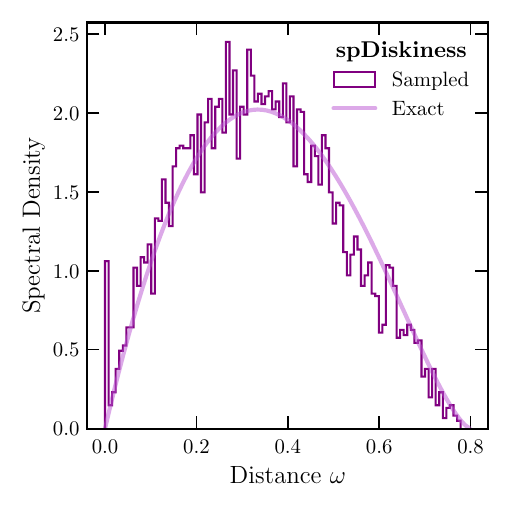}
    }
    \caption{The (a) detector representation from \Eq{disk_density} and (b) spectral representation from \Eq{disk_spectral} of an idealized disk jet, in a patch of the rapidity-azimuth plane. Each dot is a sampled particle, with its size representing its energy weight, which happens to be uniform.}
    \label{fig:example_disk}
\end{figure}

The last jet shape observable we present is the jet \emph{sDiskiness}, which asks how much like a uniform disk a given jet looks like.
The corresponding disk-like events are given by:
\begin{align}
    \E_\text{Jet Disk}^{(R)}(x) = \frac{E_{\rm tot}}{\pi R^2}\,\Theta(|x| < R), \label{eq:disk_density}
\end{align}
where $R$ is a free parameter determining the radius of the disk.
The jet sDiskiness and $R_{\rm opt}$ can be estimated using the \Specter algorithm by sampling points uniformly on a disk with radius $R$, and then performing gradient descent to minimize the $\SEMD$ with respect to $R$.
An example disk is shown in \Fig{example_disk}.

However, unlike the above cases, a closed form expression for $\O_{\rm Jet Disk}(s_\E)$ or $R_{\rm opt}(s_\E)$ seems intractable.
The spectral density of a disk is given by:\footnote{This spectral density is the solution to the well-known ``ball line picking'' problem~\cite{solomon1978geometric,klain1997introduction,santalo2004integral}. }
\begin{align}
    s_\text{Jet Disk}^{(R)}(\omega) = \frac{4\omega E_\text{tot}^2}{\pi R^2}\cos^{-1}\frac{\omega}{2R}-\frac{2\omega^2E_\text{tot}^2}{\pi R^3}\sqrt{1-\frac{\omega^2}{4R^2}}\,. \label{eq:disk_spectral}
\end{align}
To perform the closed-form $\SEMD$, one must obtain the inverse of the cumulative spectral function, and then integrate it against \Eq{simpler_inverse_cumulative_distribution}.
This is a difficult integral -- one can attempt to numerically estimate it via Monte Carlo, but this is effectively what the \Specter algorithm already does by sampling.

\subsection{Spectral Event Shapes}\label{sec:event_shapes}

We can also define spectral shape observables over the entire detector space rather than just a localized region.
The observables listed in this subsection probe the full event geometry.
For these observables, we assume that the ground metric $\omega$ is the arc-length metric on the sphere of \Eq{metric_sphere}.\footnote{One can instead choose to use other metrics, such as chord length, to define these observables. In this case, all of these observables are still well defined and qualitatively similar to the arc-length observables, but the closed-form expressions may be different (if they are derivable at all).}
All the previously described observables in \Secs{prong_shapes}{jet_shapes} can also be defined on the sphere by appropriately substituting in the correct ground metric $\omega$.

As discussed in \Sec{degeneracies}, the EMD and SEMD of localized events are expected to agree exactly when one of the events has only a single particle. 
However, when considering a full detector in the center-of-mass frame, events with a single massless particle are unphysical, and events cannot be localized to a single patch. %
Therefore, for event shapes, it is often convenient to adopt observable-dependent normalizations such that the EMD and SEMD agree on some other physical configuation.

\subsubsection{Event sThrust}\label{sec:eventthrust}

In analogy with the ordinary event thrust observable~\cite{Farhi:1977sg, BRANDT196457, DERUJULA1978387}, we can define the event \emph{sThrust}, which probes how back-to-back an event is using the SEMD.
The sThrust is given by the minimum spectral distance to a back-to-back 2 particle event with potentially different energies:
\begin{align}
    \E_{\rm Thrust}^{(z)}(x) = E_{\rm tot}\big(z \delta(\theta) + (1-z)\delta(\theta - \pi)\big), \label{eq:thrust_density}
\end{align}
where $z$ and $1-z$ parameterize the energy fractions carried by each particle and $\theta$ is the usual polar coordinate of the sphere.
The corresponding spectral function is
\begin{align}
s_\text{\rm Thrust}^{(z)}(\omega) = \left(
z^2+(1-z)^2
\right)E_\text{tot}^2\delta(\omega) + 2z(1-z)E_\text{tot}^2\delta(\omega - \pi)\,.\label{eq:thrust_spectral}
\end{align}

\begin{figure}[t!]
    \centering
    \subfloat[]{
        \includegraphics[width=0.6\textwidth]{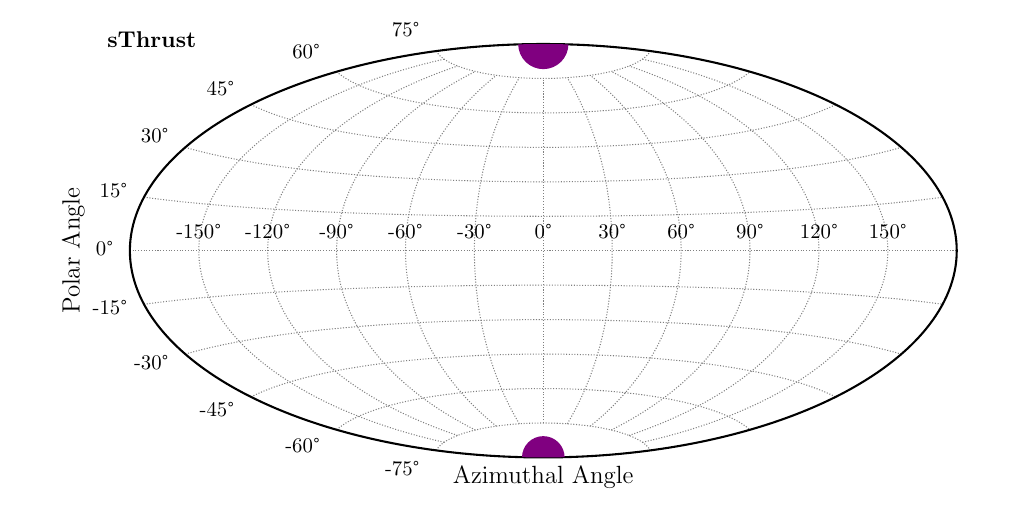}
    }
    \subfloat[]{
        \includegraphics[width=0.35\textwidth]{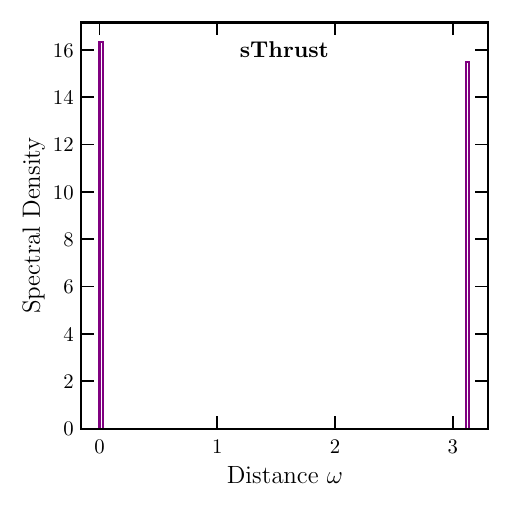}
    }
    \caption{The (a) detector representation from \Eq{thrust_density} and (b) spectral representation from \Eq{thrust_spectral} of an idealized thrust event with $z \approx 0.45$, on the celestial sphere. Each dot is a particle, with its size representing its energy weight.}
    \label{fig:example_thrust}
\end{figure}

Note that unlike the case with ordinary Thrust, which when calculated using the EMD requires optimizing over both $z$ and a thrust axis, rotational symmetry removes the need optimize over a thrust axis for sThrust, and thus only a single optimization over $z$ is required.
The sThrust is actually just a special case of the event 2-sPronginess discussed in \Sec{n_spronginess}, with the particle positions fixed. 
An example of a thrust configuration is shown in \Fig{example_thrust}.

The event sThrust of an event $\E$ is then given by:
\begin{align}
    \O_{\rm Thrust}(s_\E) = \frac{2}{\pi^2}\min_{z\in [0,1]} \SEMD(s_E, s_{\rm Thrust}^{(z)}),
\end{align}
with the factor of $\frac{2}{\pi^2}$ chosen such that the sThrust is normalized between 0 and $\frac{1}{2}$, like the ordinary Thrust.\footnote{In our convention, sThrust is low for pencil-like events and large for events away from pencil-like. This differs from the ``\FastJet'' Thrust, which does the opposite.
The sThrust really corresponds to $(1-\text{Thrust})$.}
This normalization reflects the choice of using arc-length rather than chord length; the ordinary thrust is typically defined through dot products between the particle momentum and the thrust axis (resulting in a chord length) and is thus slightly different, but in the collinear limit these are linearly related by this normalization factor.
One can, of course, have used chord lengths to define sThrust instead in closer correspondence with the ordinary Thrust, which we discuss in \App{thrust_chord}.

The SEMD between the thrust event and a general event ${\cal E}$ is
\begin{align}
\text{SEMD} &=\min_{z\in[0,1]}\sum_{i<j\in {\cal E}}2E_iE_j\omega_{ij}^2+2\pi^2z(1-z)\,E_\text{tot}^2 \\
&\hspace{1cm}-2\pi\sum_{\substack{n\in {\cal E}^2\\ \omega_n<\omega_{n+1}}}\omega_n \,\text{ReLU}\left(
 S^+(\omega_n)-\max\left[\left(
z^2+(1-z)^2
\right)E_\text{tot}^2, S^-(\omega_n)\right]
\right).\nonumber
\end{align}
The optimization over $z$ can be performed via gradient descent. 
However, it can be shown that $z = \frac{1}{2}$ is a local extremum of the SEMD (see \App{thrust_chord} for more details), and is thus not an ideal initial point for gradient descent.
The value $z = \frac{1}{2}$ is not necessarily the true minimum, merely an extremum, though in practice we find that the true minimum is extremely close to $z = \frac{1}{2}$ in dijet events.
It may also be possible to perform the minimization \emph{exactly} in finite time as a distribution-partitioning problem in spectral space, analogous to the hemisphere-picking problem for ordinary Thrust (see \Reference{Wei:2019rqy} for a review of classical and quantum algorithms), though we do not pursue this further here.
There is also a simple test for checking if $z = 0.5$ is the true minimum, given by  \Eq{thrust_check} in \App{thrust_chord}.

\subsubsection{Event spIsotropy}\label{sec:eventisotropy}

\begin{figure}[t!]
    \centering
    \subfloat[]{
        \includegraphics[width=0.6\textwidth]{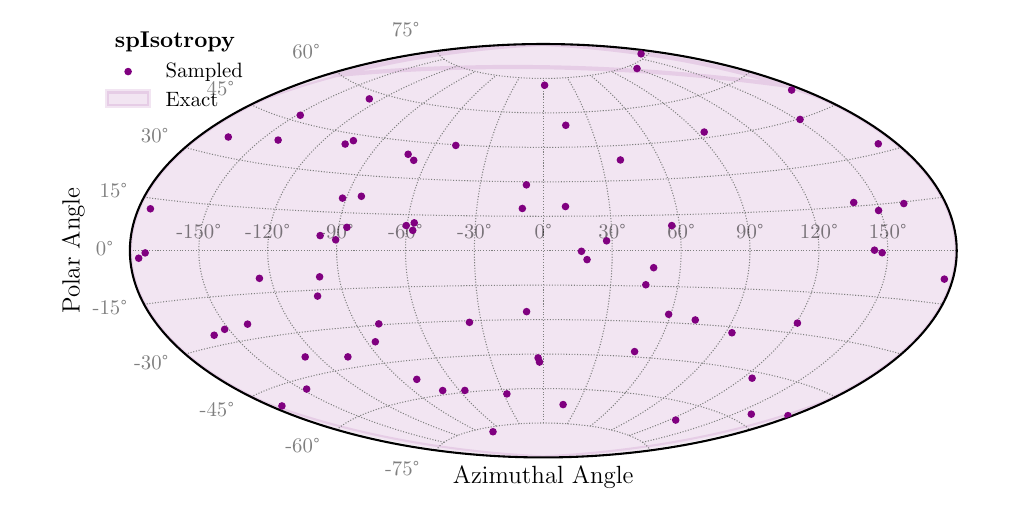}
    }
    \subfloat[]{
        \includegraphics[width=0.35\textwidth]{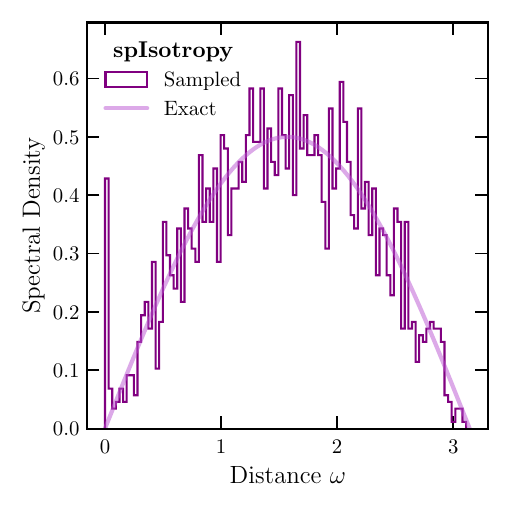}
    }
    \caption{The (a) detector representation from \Eq{isotropy_density} and (b) spectral representation from\Eq{isotropy_spectral} of an idealized isotropic event, on the celestial sphere. Each dot is a sampled particle, with its size representing its energy weight, which happens to be uniform.}
    \label{fig:example_isotropy}
\end{figure}

Next, we define the spectral Isotropy (\emph{spIsotropy}) of an event over the celestial sphere.
In analogy with the ordinary Isotropy~\cite{Cesarotti:2020hwb}, this is defined as the (spectral) distance to the uniform event over the sphere:
\begin{align}
    \E_{\rm Isotropy}(x) = \frac{E_{\rm tot}}{4\pi} \label{eq:isotropy_density}.
\end{align}
The corresponding spectral representation can be calculated to be:
\begin{align}
    s_{\rm Isotropy}(\omega) = \frac{E_{\rm tot}^2}{2}\sin(\omega).\label{eq:isotropy_spectral}
\end{align}
Like the continuous observables described in the previous subsection, the uniform event can be approximated by sampling $N$ points on the sphere, each with energy $\frac{E_{\rm tot}}{N}$.
Then, all one has to do is use the \Specter algorithm to compute the $\SEMD$ between the event of interest $\E$ and this discrete event.
There are no free parameters in \Eq{isotropy_density}, so no minimization is necessary. 
An example of such a sampling is shown in \Fig{example_isotropy}.

Like the event sThrust, we find it convenient to normalize the event spIsotropy. 
By convention, we multiply the SEMD value by a factor of $\frac{1}{\pi-2}$, such that the event spIsotropy of a perfectly back-to-back equal energy event is 1, to match the ordinary spherical Isotropy.

The spIsotropy observable can be written exactly in closed-form, allowing one to bypass the \Specter algorithm for generic observables.
By using \Eq{semd2start} and the exact form of the cumulative spectral function for a uniform sphere (see \App{closed_form_spisotropy} for details), one can show:
\begin{align}\label{eq:spisotropyobs}
    \O_{\rm Isotropy}(s_\E) &= \frac{1}{\pi-2}\left(\sum_{i<j\in{\cal E}}2E_iE_j\omega_{ij}^2 + \frac{\pi^2 - 4}{2}E_{\rm tot}^2 - 2\sum_{\substack{n\in{\cal E}^2\\\omega_n<\omega_{n+1}}}\omega_n \left[f^{+}(n) - f^{-}(n)\right]\right),
\end{align}
where
\begin{align}
    f^{\pm}(n) &= \sqrt{S^{\pm}(\omega_n)}\sqrt{E^{2}_{\rm tot} - S^{\pm}(\omega_n)} + S^{\pm}(\omega_n) \cos^{-1}\left(1 - 2\frac{S^{\pm}(\omega_n)}{E_{\rm tot}^2}\right) \\
    &\hspace{7cm}-\, E_{\rm tot}^2 \sin^{-1}\left(\frac{\sqrt{ S^{\pm}(\omega_n)}}{E_{\rm tot}}\right)\,.\nonumber
\end{align}

\subsubsection{Event spEquatorialness}\label{sec:eventringiness}

\begin{figure}[t!]
    \centering
    \subfloat[]{
        \includegraphics[width=0.6\textwidth]{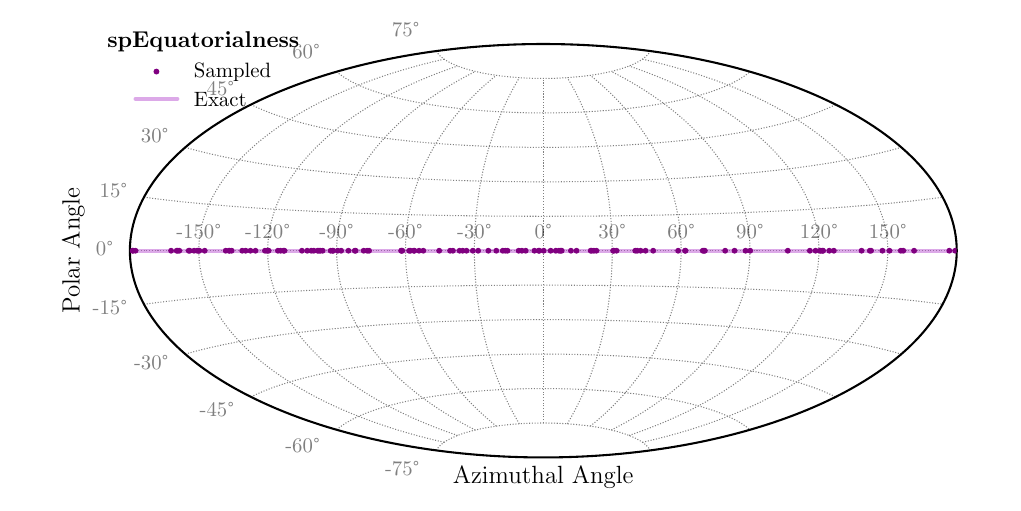}
    }
    \subfloat[]{
        \includegraphics[width=0.35\textwidth]{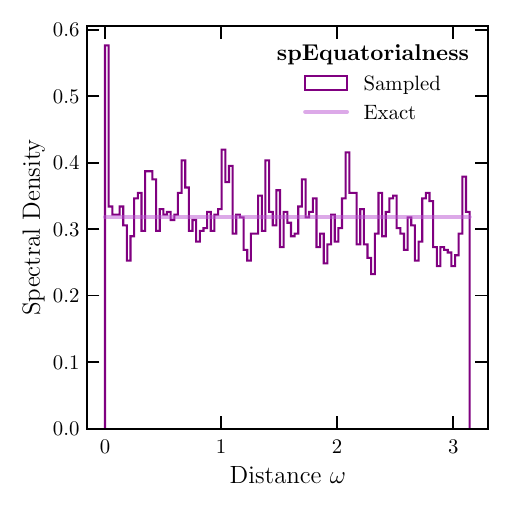}
    }
    \caption{The (a) detector representation from \Eq{eringiness_density} and (b) spectral representation from \Eq{eringiness_spectral} of an idealized ring event, on the celestial sphere. Each dot is a sampled particle, with its size representing its energy weight, which happens to be uniform.}
    \label{fig:example_ering}
\end{figure}

On the celestial sphere, we can define the event \emph{spEquatorialness} as the distance of an event to the ring-like event on the equator of the sphere, given by:
\begin{align}
    \E_{\rm Equator}(x) = \frac{E_{\rm tot}}{2\pi}\, \delta\left(\theta - \frac{\pi}{2}\right), \label{eq:eringiness_density}
\end{align}
where $\theta$ is the usual polar coordinate of the sphere.
Due to rotational invariance, all other great circles on the sphere are the same in spectral space.
The corresponding spectral representation is:
\begin{align}
    s_{\rm Equator}(\omega) = \frac{E_{\rm tot}^2}{\pi}. \label{eq:eringiness_spectral}
\end{align}
    Just like the event spIsotropy, the event spEquatorialness can be easily estimated by sampling $N$ points uniformly on the equator of the sphere using the \Specter algorithm.
An example of this sampling is shown in \Fig{example_ering}.

However, it is also possible to solve for the $\SEMD$ exactly in closed form by using \Eq{semd2start} and the cumulative spectral function for a ring (see \App{closed_form_event_springiness} for details):
\begin{align}\label{eq:eventringobs}
    \O_{\rm Equator}(s_\E) &= \sum_{i<j\in {\cal E}}2E_iE_j\omega_{ij}^2+\frac{\pi^2}{3}\,E_\text{tot}^2\\
&\hspace{2cm}-\frac{\pi}{E_\text{tot}^2}\sum_{\substack{n\in{\cal E}^2\\\omega_n<\omega_{n+1}}}\omega_n\left[
\left(
S^+(\omega_n)
\right)^2-\left(
S^-(\omega_n)
\right)^2
\right]\,.\nonumber
\end{align}

\subsection{...And More!}

It is important to emphasize that the above list of shape observables is \emph{not} exhaustive. 
As with \Shaper, one can use \emph{any} parameterized manifold of events $\E^{(\theta)}$ to define a shape, so long as it is known how to sample from it.
This includes more exotic shapes such as ellipses or polygons, but also combinations of shapes such as $N$-spRinginess, or $N$-sPronginess plus spIsotropy defined as $N$ prongs on top of a uniform background.
All of these are possible within the \Specter framework.

Many of the observables discussed above were simple enough to admit an exact closed-form expression, bypassing the need to use the \Specter algorithm at all to estimate them.
However, this is not the case for a generic spectral observable -- for these, one must use \Specter to estimate them.

\section{Empirical Studies}\label{sec:compstudies}

In this section, we use \Specter to evaluate $\SEMD$s and spectral shape observables on simulated collider events.
We begin with a comparison of pairwise (S)EMDs between jets and between full events, to explore the differences between the SEMD and ordinary EMD.
Then, we evaluate the spectral shape observables defined in \Sec{specshapes} on both localized jets and global events. 
These are evaluated using the numeric \Specter algorithm, and, wherever possible, also in closed-form.
For comparison, we show the distribution of the corresponding ordinary EMD-based shapes, as defined in \Reference{Ba:2023hix} using the \Shaper algorithm for $\beta = 2$.\footnote{To match conventions, we multiply the ordinary EMD as evaluated by \Shaper by a factor of 2, since the metric used in \Reference{Ba:2023hix} is $\frac{1}{\beta}|x-y|^{\beta/2}$ rather than the $|x-y|^{\beta/2}$ used here.  This factor was chosen in \Reference{Ba:2023hix} so that the gradients of the Sinkhorn divergence have nicer scaling with $\beta$ and to match onto other metrics in the optimal transport literature.}

\subsection{Datasets}\label{sec:datasets}

For the empirical studies in this section, we generate several different sample datasets using $\textsc{Pythia}$ 8.3 \cite{Bierlich:2022pfr}. 
The first three datasets consist of localized  jets for studying the properties of pairwise (S)EMDs and jet shape observables.
The final, fourth dataset consists of full events for the purposes of studying event shape observables.

\begin{figure}[tpb!]
    \centering
    \subfloat[]{
        \includegraphics[width=0.46\textwidth]{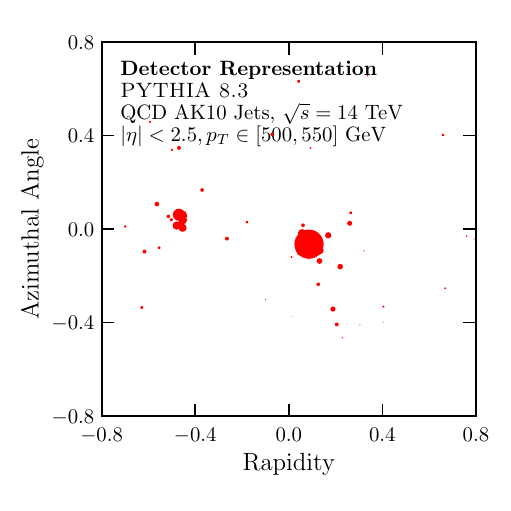}
    }
    \subfloat[]{
        \includegraphics[width=0.45\textwidth]{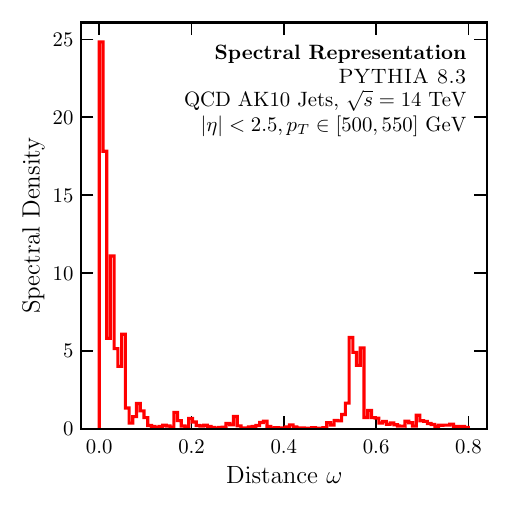}
    }
    \caption{The (a) detector representation and (b) spectral representation of a typical simulated QCD jet, in a patch of the rapidity-azimuth plane.}
    \label{fig:example_event}
\end{figure}

\begin{figure}[tpb!]
    \centering
    \subfloat[]{
        \includegraphics[width=0.6\textwidth]{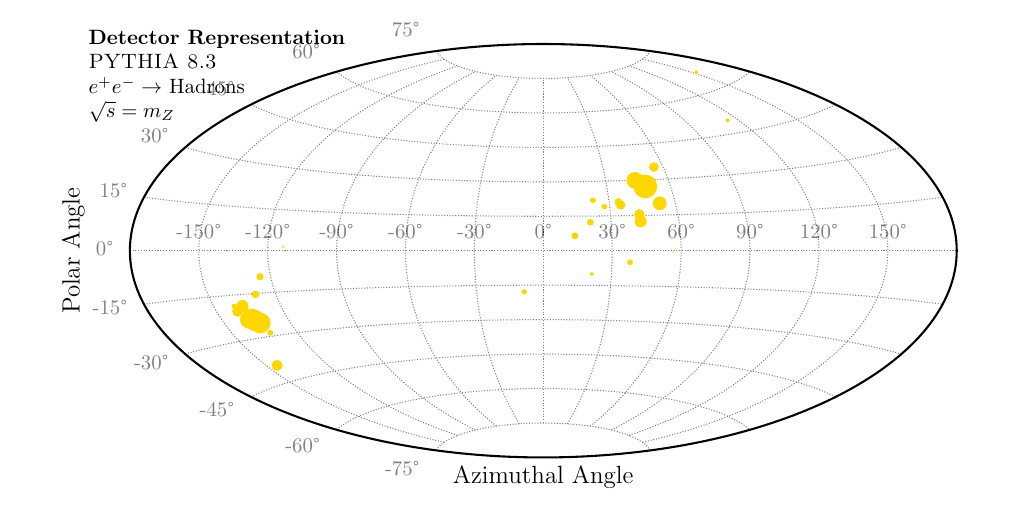}
    }
    \subfloat[]{
        \includegraphics[width=0.35\textwidth]{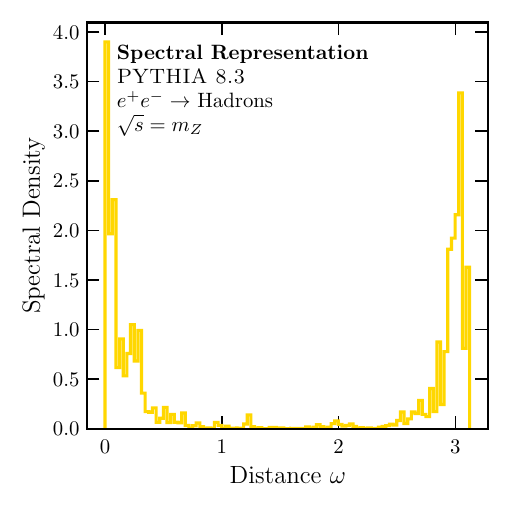}
    }
    \caption{The (a) detector representation and (b) spectral representation of a typical simulated $e^+ e^- \to \text{hadrons}$ event, on the celestial sphere.}
    \label{fig:example_sphere}
\end{figure}

For each dataset described below, we use 100k samples, with this number chosen because \Shaper (but not \Specter) is often time-limited.
All datasets are publicly available, as described in \hyperref[ref:code_data]{Code and Data} below.
The datasets are:

\begin{itemize}
    \item \textbf{QCD Jets:} We generate QCD events in proton-proton collisons at $\sqrt{s} = 14$ TeV using $\textsc{Pythia}$'s $\texttt{HardQCD:all}$ process, with a phase space cut of $\hat{p}_T \in [375, 687.5]$ GeV. 
    Multi-parton interactions, initial-state radiation, and final-state radiation are kept on.
    The events are then clustered into jets using \FastJet 3.3.0 \cite{Cacciari:2011ma} using the anti-$k_T$ algorithm \cite{Cacciari:2008gp} with jet radius $R = 1.0$ (AK10 ``fat jets'').
    Final-state invisible particles, namely neutrinos, are not included in the clustering.
    Jets are required to satisfy $p_{T,J} \in [500, 550]$ GeV and $\eta_J < |2.5|$.
    The jets are then saved. 
    If an event has multiple jets meeting all cuts, only one jet is saved, selected randomly.
    The 125 most energetic particles are saved per jet.
    These jets are used for the pairwise (S)EMD studies in \Sec{empirical_pairwise} and for the jet shape studies in \Sec{empirical_jet_shapes}.
    An example of a typical jet in this dataset, along with its spectral representation, is shown in \Fig{example_event}.
    \item \textbf{Top Quark Jets:} We generate $\Bar{t}t$ events in proton-proton collisons  at $\sqrt{s} = 14$ TeV using $\textsc{Pythia}$'s $\texttt{Top:gg2ttbar}$ and $\texttt{Top:qqbar2ttbar}$ processes, with a phase space cut of $\hat{p}_T \in [375, 687.5]$ GeV. 
    Multi-parton interactions, initial-state radiation, and final-state radiation are kept on.
    All tops are forced to decay hadronically.
    Events are then clustered into AK10 jets, exactly as in the QCD jets above.
    Jets are required to satisfy $p_{T,J} \in [500, 550]$ GeV and $\eta_J < |2.5|$.
    Additionally, we require that there exists a top quark parton within $\Delta R = 1.0$ of the jet axis.
    One jet amongst those that satisfy the cuts is randomly selected per event to be saved as a $t$ jet. 
   The 125 most energetic particles are saved per jet.
    The $t$ jets are used for the pairwise (s)EMD studies in \Sec{empirical_pairwise}.
    \item \textbf{Uniform Phase Space ``Jets'':} We generate points in uniform phase space using the RAMBO algorithm \cite{Kleiss:1985gy}.
    Each event has 125 particles with at total $\sqrt{s} = 150$ GeV.
    These events are then boosted in the $x$-direction such that their total energy is 500 GeV, which makes the events appear jet-like, but without the singularity structure of QCD.
    These ``jets'' are used for the pairwise (S)EMD studies in \Sec{empirical_pairwise}.
    \item \textbf{LEP Dijet Events:}
    We generate $e^+e^- \to \text{hadrons}$ events at $\sqrt{s} = m_Z$ using $\textsc{Pythia}$'s $\texttt{WeakSingleBoson:ffbar2gmZ}$ process, in order to simulate LEP-like jets.
    The intermediate virtual photon or $Z$ is forced to decay to one of the 5 light quarks.
    Initial-state radiation due to the leptons is turned off.
    No additional cuts or clustering are applied.
    An example of a typical event in this dataset, along with its spectral representation, is shown in \Fig{example_sphere}.
\end{itemize}

\subsection{Pairwise Distances between Events and Jets}\label{sec:empirical_pairwise}

In this subsection, we compute the (spectral) EMD between pairs of jets associated with different underlying processes.
As discussed in \Sec{degeneracies}, one should not generally expect the $\SEMD$ and ordinary $\EMD$ to be the same, though they may be correlated. 
Whenever an (almost) equilateral triangle configuration occurs between any set of three particles in an event, one should expect a deviation between the SEMD and EMD, as the information probed is genuinely different.
These configurations occur whenever the energies of particles are all within an $\O(1)$ number of each other, and their pairwise distances are also all within an $\O(1)$ number of each other.
There are two factors we can identify that lead to equilateral triangle configurations:
\begin{enumerate}
    \item \textbf{Number of Particles:} Combinatorially, if an event has $N$ particles, we can form ${N \choose 3}$ triangles between them. 
    If the energies and positions of particles are uniformly distributed, one should expect that the number of equilateral triangles should scale similarly.
    In QCD jets, however, emissions off the hard core are strongly ordered in both energy and angle, so forming equilateral triangles with balanced energies involving the hard core is highly suppressed.
    \item \textbf{Scaling:}  Even if there do exist equilateral triangle configurations, if the pairwise $(2E_iE_j)$ or the $\omega_{ij}$ scales are small, then the degeneracy will contribute very little to the EMD/SEMD difference. 
    In QCD jets, these are the soft and collinear scales, which are suppressed.
    In top jets, there is an additional hard scale, $m_t / E$, to contend with.
\end{enumerate}
Thus, we should expect that as we increase the number of particles in an event, or as we change the event type to include harder-scale physics, that the EMD and SEMD become more different.
To show this, we calculate pairwise SEMDs (using \Specter) and isometry-modded EMDs (using \Shaper) between QCD jets, top quark jets, and uniform phase space, each described in \Sec{datasets}.
Moving from QCD to top jets to uniform phase space, one should expect that the emissions are less strongly-ordered and that there exist harder and wider scales to form equilateral triangles.
Each dataset is clustered  into 2, 5, and 25 subjets with the exclusive $k_T$-algorithm~\cite{Catani:1993hr, Ellis:1993tq} in \FastJet~\cite{Cacciari:2011ma}.

\begin{figure}[tpb!]
    \centering
        \includegraphics[width = 1.0\linewidth]{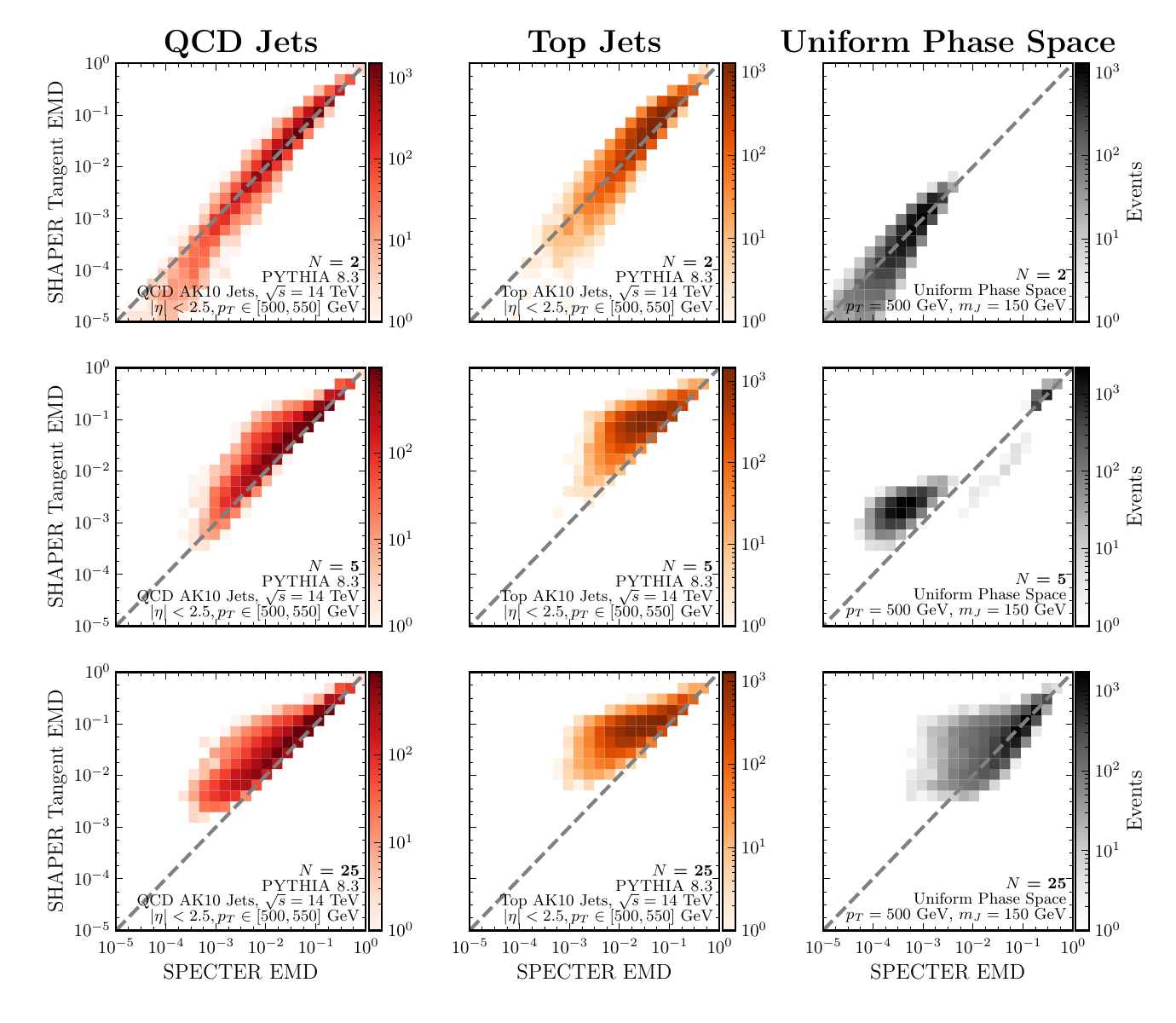}
        \caption{Comparison of the $\SEMD$, as evaluated using \Specter, to the isometry-modded $\EMD$, as evaluated using \Shaper, on pairs of events. 
        Pairs of QCD jets are used for the left column (red), pairs of top jets are used for the middle column (orange), and pairs of uniform boosted phase space are used for the right column (yellow). Each jet is clustered into subjets before the (S)EMD is computed: 2 subjets in the top row, 5 subjets in the middle row, and 25 subjets in the bottom row.
        Note that the histogram is in log scale.}
    \label{fig:comparison}
\end{figure}

Unlike the $\SEMD$, the ordinary $\EMD$ is not invariant under isometries. 
To alleviate this and put our comparisons on fairer ground, we use the ``Tangent EMD''~\cite{10.1007/978-3-642-40020-9_43}, wherein we ``mod out'' the isometries by minimizing the $\EMD$ over all possible translations and rotations using \Shaper. 
Precisely, this is given by:
\begin{align}
    \text{Tangent EMD}(\E, \E') = \min_{\theta \in E(2)}\EMD(\E, \E'^{(\theta)}),
\end{align}
where $E(2)$ is the two-dimensional Euclidean group, and $\E'^{(\theta)}$ is a translated and rotated energy flow parameterized by $\theta$.

The results of this study are shown in \Fig{comparison}.
In the grid of plots, $N$ is increased going downwards and harder physics is shown going rightwards.
In both directions, the SEMD to EMD difference grows relative to the baseline $N = 2$ QCD jets in the upper left. 
Moreover, the EMD is often greater than the SEMD, but this is not always the case.
We can make several observations from this plot:
\begin{itemize}
    \item \textbf{SEMD $\approx$ EMD in QCD jets:} Even across a large number of $k_T$ subjets, the EMD and SEMD are very close, especially compared to top jets or uniform phase space. This follows from the above arguments -- there are typically no additional hard emissions in QCD jets that can ``spoil'' the spectral representation with a near-degeneracy.
    \item \textbf{SEMD $\lesssim$ EMD generically:} It seems there is a slight bias for the EMD to be greater than the SEMD rather than the other way around. It is hard to tell if this is a result of imperfect minimization while calculating the Tangent EMD, or a robust feature -- one might expect that the existence of near-degeneracies would generically decrease distances in spectral space, though Sec 7.3 of \Reference{Larkoski:2023qnv} provides an example where the SEMD can be greater.
    \item \textbf{Differences grow with $N_{\rm subjets}$:} The relative SEMD to EMD difference of the distributions grows as the number of subjets increase. In particular, for $N = 2$, the two are highly correlated across all three types of jets. 
\end{itemize}

There is clear evidence of a physics-dependent correlation between the EMD and SEMD, potentially intertwined with the collinear and strongly-ordered structure of QCD. 
There is more to be done to elucidate the relationship between these quantities, which we leave for future work.

\subsection{Hearing Jets spRing and sProng}\label{sec:empirical_jet_shapes}

\begin{figure}[t]
    \centering
        \subfloat[]{    \includegraphics[width=0.42\textwidth]{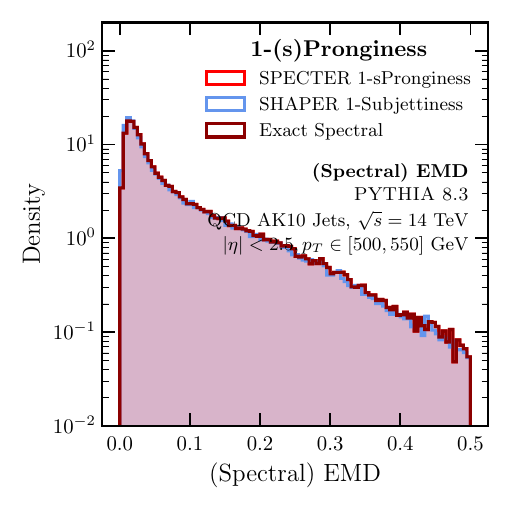}
        }\label{fig:1sprong_emd}
        \subfloat[]{    \includegraphics[width=0.42\textwidth]{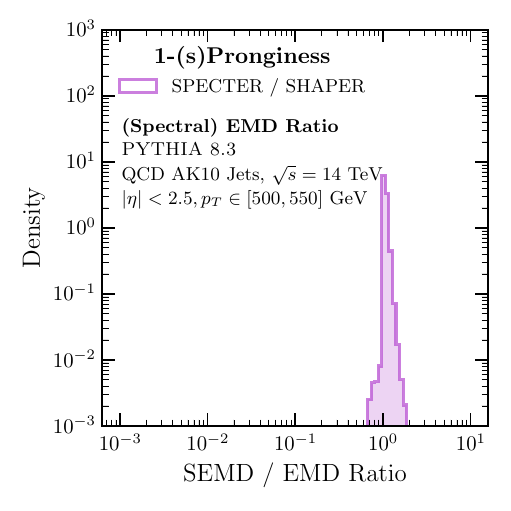}
        }\label{fig:1sprong_ratio}
        \caption{(a) Distributions of the 1-(s)Pronginess observable, as defined in \Sec{1_spronginess}, evaluated on jets using \Specter (red) and \Shaper (purple). (b) The corresponding ratio of the sPronginess to the Pronginess.}
    \label{fig:1pronginess}
\end{figure}
In this subsection, we evaluate the jet shape observables defined in \Sec{specshapes} on the QCD jets dataset described in \Sec{datasets}.
For each observable, we evaluate both the spectral shape observable $\mathcal{O}_{\rm Shape}(s_\E)$ and the corresponding spectral shape parameters $\theta_{\rm Shape}(s_\E)$ on the distribution of events $\E$ using \Specter.
We also show the distributions for the corresponding EMD-based observables defined using the same shapes, as evaluated using \Shaper.

For both programs, we sample $N = 125$ points for each shape. 
The gradient descent operation is run for 500 epochs with the Adam optimizer with a learning rate of $0.001$. 
Often, convergence is reached by $\sim$100 epochs, but we do not do any early stopping to be conservative.
\Shaper requires two additional parameters that control the Sinkhorn estimation: we choose $\epsilon = 10^{-3}$ and $\Delta = 0.99$ for results as precise as possible.

\begin{figure}[p!]
    \centering
        \subfloat[]{    \includegraphics[width=0.42\textwidth]{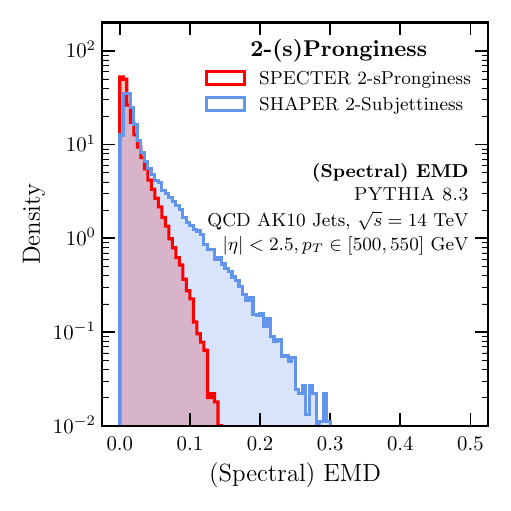}
        }\label{fig:2sprong_emd}
        \subfloat[]{    \includegraphics[width=0.42\textwidth]{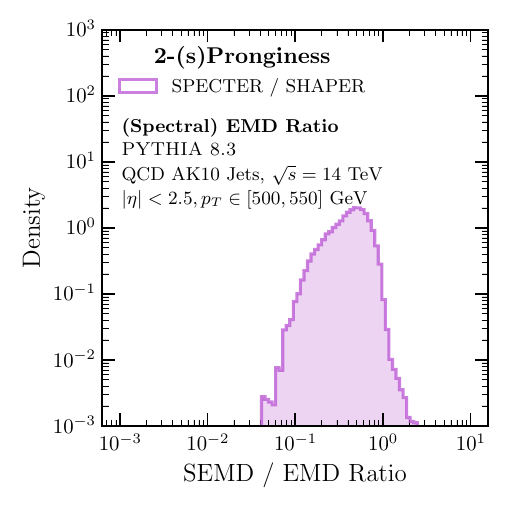}
        }\label{fig:2sprong_ratio}
        \vspace{-10pt}
        \subfloat[]{    \includegraphics[width=0.42\textwidth]{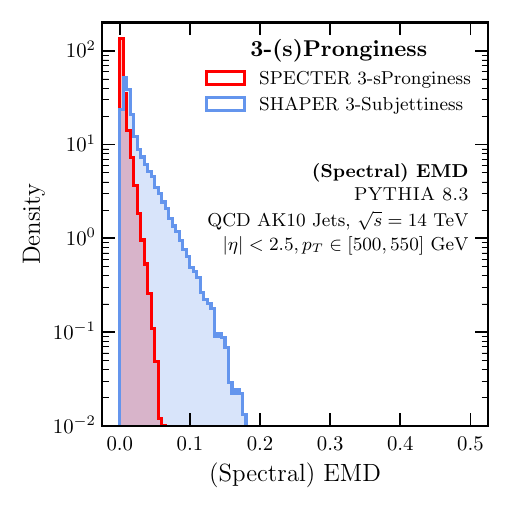}
        }\label{fig:3sprong_emd}
        \subfloat[]{    \includegraphics[width=0.42\textwidth]{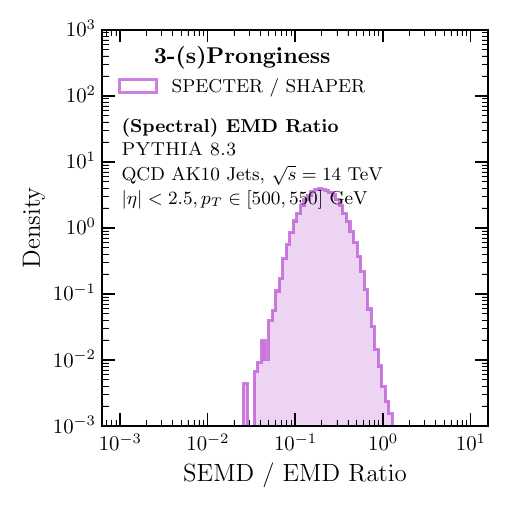}
        }\label{fig:3sprong_ratio}
        \caption{Distributions of the 2- and 3- (s)Pronginess observables, as defined in \Sec{n_spronginess}, evaluated on jets using \Specter (red) and \Shaper (purple) in (a) and (c), respectively. The corresponding ratio of the sPronginess to the Pronginess is shown in (b) and (d), respectively.}
    \label{fig:23pronginess}
\end{figure}

In \Figs{1pronginess}{23pronginess}, we show the 1-(s)Pronginess, 2-(s)Pronginess, and 3-(s)Pronginess, as defined in \Secs{1_spronginess}{n_spronginess}.
The spectral 1-sPronginess and the ordinary 1-Pronginess (1-subjettiness) should be identical -- indeed, we see that the ratio of the two is close to 1. 
Deviations from 1 are due to numerics: both \Specter and \Shaper use a finite gradient descent algorithm (though \Specter's gradient descent happens to be irrelevant for 1-sPronginess), and \Shaper uses an approximation of the EMD rather than the exact EMD.
The spectral 2- and 3-sPronginess, on the other hand, are significantly less than their ordinary EMD counterparts, and are less correlated.

\begin{figure}[p]
\centering
\subfloat[]{
\includegraphics[width=0.45\textwidth]{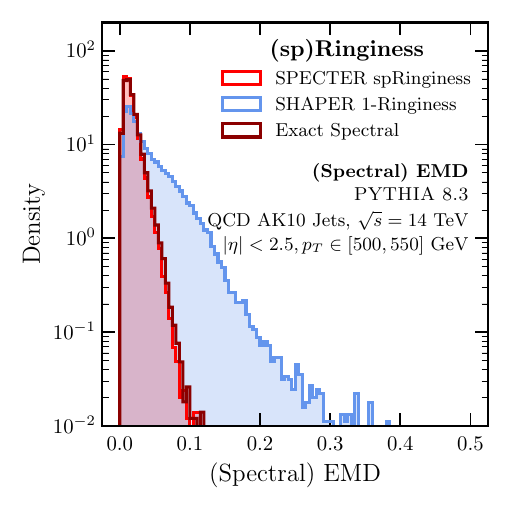}
}\label{fig:ring_emd}
\subfloat[]{
\includegraphics[width=0.45\textwidth]{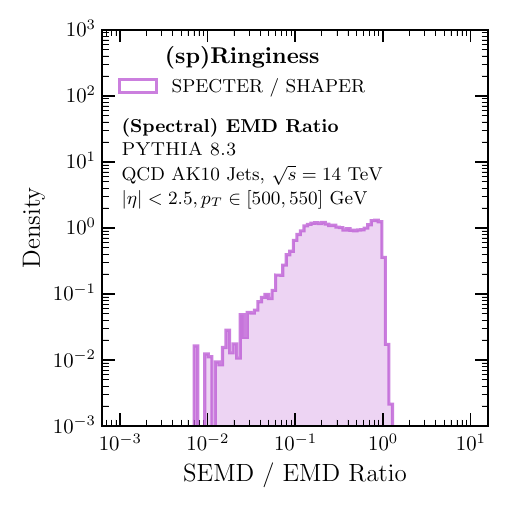}
}\label{fig:ring_ratio}
\vspace{1em}
\subfloat[]{
\includegraphics[width=0.45\textwidth]{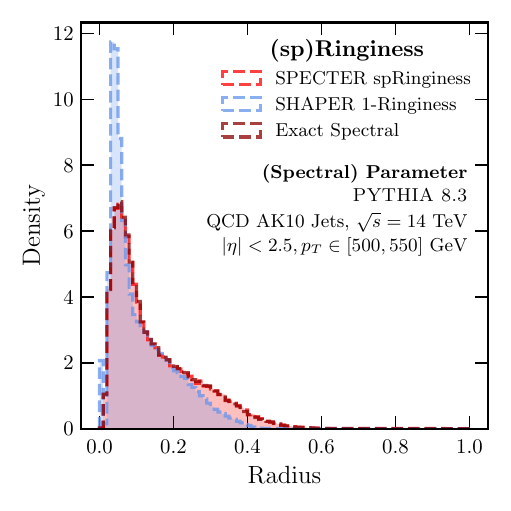}
}\label{fig:ring_param}
\subfloat[]{
\includegraphics[width=0.45\textwidth]{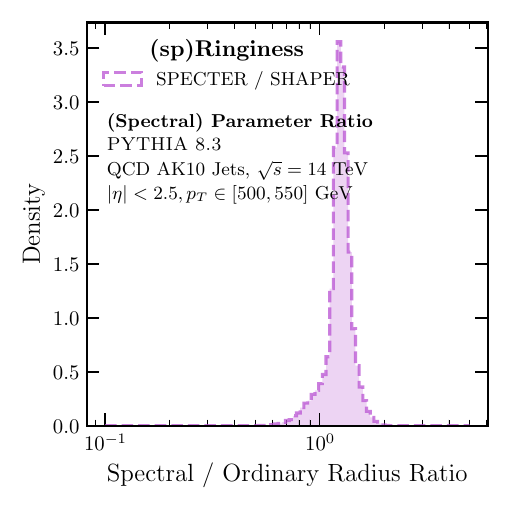}
}\label{fig:ring_param_ratio}
\caption{Distributions for the (sp)Ringiness observable, as defined in \Sec{jetringiness}, evaluated on jets using \Specter (red) and \Shaper (purple). The observable value is shown in (a), and the corresponding shape parameter is shown in (c). The ratio of the spRinginess to the Ringiness is shown in (b) and (d).}
\label{fig:rings}
\end{figure}

In \Figss{rings}{lines}{disks}, we show the jet (sp)Ringiness, the jet (sp)Lineliness, and the jet (s)Diskiness respectively, as defined in \Sec{jet_shapes}.
Each of these observables is defined with a corresponding length scale parameter (radii for the rings and disks, and lengths for the lines).
For each observable, we show the (spectral) EMD (solid) as well as the corresponding length scale parameter (dashed) as calculated using \Specter and \Shaper.
In all cases, the spectral observable is smaller than the corresponding ordinary observable by factor of about $2$ to $5$.
With the exception of the $1$-sPronginess, which is expected to exactly match the $1$-subjettiness, this ratio is dataset dependent and difficult to predict. 
For instance, one can construct a dataset of near-equilateral triangles, in which the 2-subjettiness is infinitely larger than the 2-sPronginess, as in \Fig{degeneracy_plot}. 
Understanding the origin of this factor of 2 to 5 times would require a full QCD computation, which we leave for future work.
Interestingly, the two circular observables achieve essentially the same radius using either the spectral or ordinary metric, but the spLineliness tends to have a length parameter 2 times bigger than the ordinary Lineliness. 
Notably, all three parameter ratios peak sharply near either 1 or 2, whereas the corresponding (spectral) EMD ratios are broad -- changing the metric seems to scale the physical distance between events, which is not unexpected, but does not seem to change the optimal geometry by much.

\begin{figure}[p]
\centering
\subfloat[]{
\includegraphics[width=0.45\textwidth]{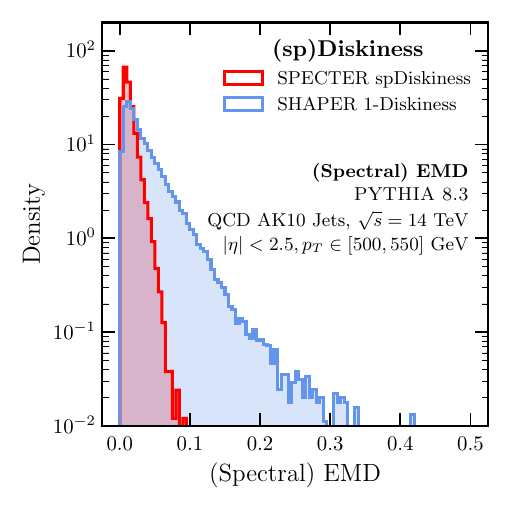}
}\label{fig:disk_emd}
\subfloat[]{
\includegraphics[width=0.45\textwidth]{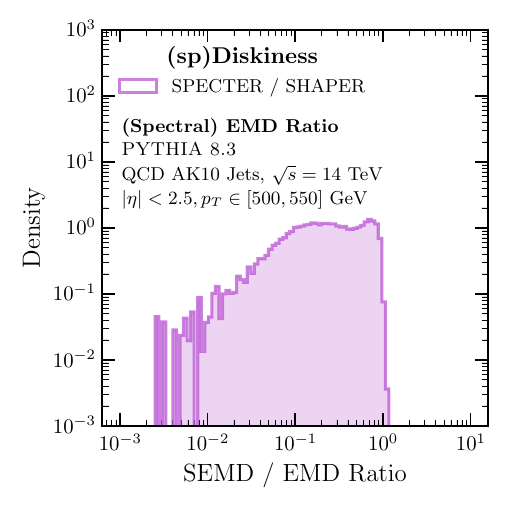}
}\label{fig:disk_ratio}
\vspace{1em}
\subfloat[]{
\includegraphics[width=0.45\textwidth]{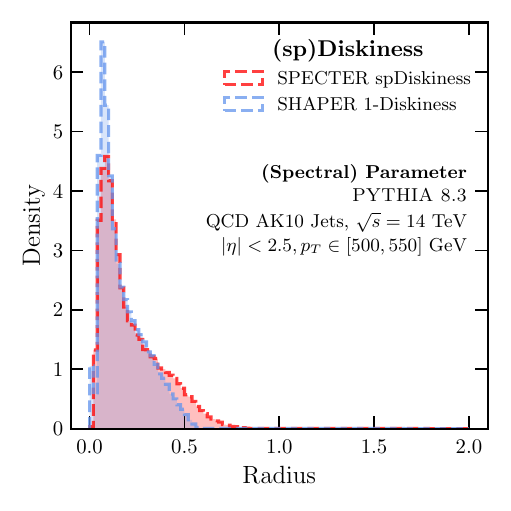}
}\label{fig:disk_param}
\subfloat[]{
\includegraphics[width=0.45\textwidth]{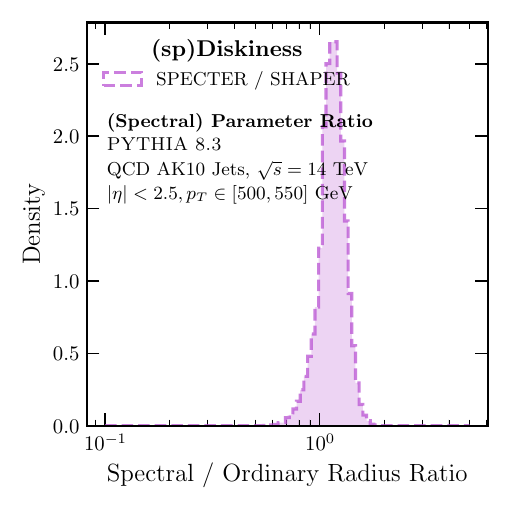}
}\label{fig:disk_param_ratio}
\caption{Distributions for the (sp)Diskiness observable, as defined in \Sec{jetdiskiness}, evaluated on jets using \Specter (red) and \Shaper (purple). The observable value is shown in (a), and the corresponding shape parameter is shown in (c). The ratio of the spDiskiness to the Diskiness is shown in (b) and (d).}
\label{fig:disks}
\end{figure}

\subsection{Event Shapes}\label{sec:empirical_event_shapes}

In this subsection, we evaluate the event shape observables defined in \Sec{specshapes} on the $e^+e^-$ dataset described in \Sec{datasets}, namely the event sThrust, event spIsotropy, and event spRinginess. 

\begin{figure}[t!]
\centering
\subfloat[]{
\includegraphics[width=0.45\textwidth]{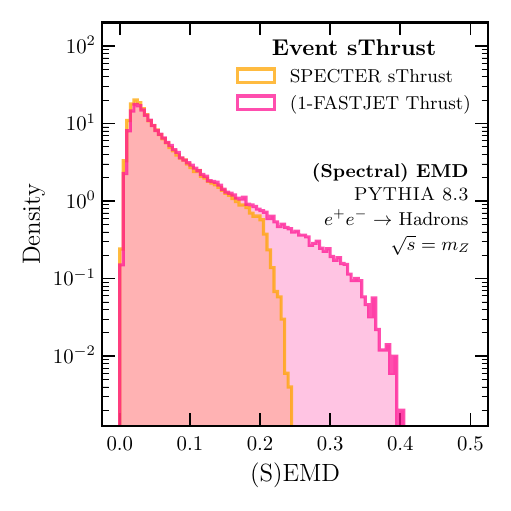}
}\label{fig:thrust_emd}
\subfloat[]{
\includegraphics[width=0.45\textwidth]{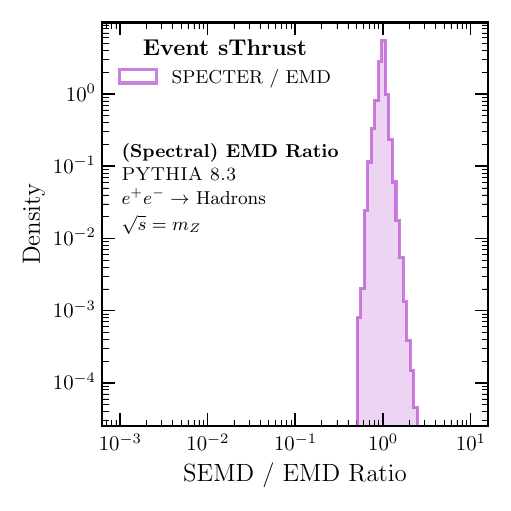}
}\label{fig:thrust_ratio}
\caption{Distributions for the event (s)Thrust observable, as defined in \Sec{eventthrust}, evaluated on jets using \Specter (orange), and the \FastJet 3.3.0 (pink). The observable value is shown in (a) and the ratio of the sThrust to the Thrust is shown in (b). A variant of the sThrust based on chord lengths rather than arc lengths is shown in \Fig{thrust_chord}.}
\label{fig:thrust}
\end{figure}

In \Fig{thrust}, we show the sThrust, as computed by the \Specter algorithm, and $(1-\text{Thrust})$, as calculated using \FastJet $3.3.0$.
The two distributions agree extremely well near (S)EMD values of 0, with the distribution of ratios sharply peaked at 1, indicating that the sThrust and Thrust behave similarly in the collinear limit, as expected. 
At larger values of the (S)EMD, however, the two distributions diverge, and the SEMD does not reach these higher values.
This is a consequence of the fact that the sThrust and Thrust use different ground metrics.
The sThrust uses the arc-length metric of \Eq{metric_sphere}, but ordinary Thrust corresponds to the chord-length metric, $\omega^2 \sim 1 - \cos(\theta)$.
If one were to define sThrust using this metric as well, the sThrust and Thrust distributions become nearly identical, as we explore further in \App{thrust_chord}. 

\begin{figure}[tpb!]
\centering
\subfloat[]{
\includegraphics[width=0.45\textwidth]{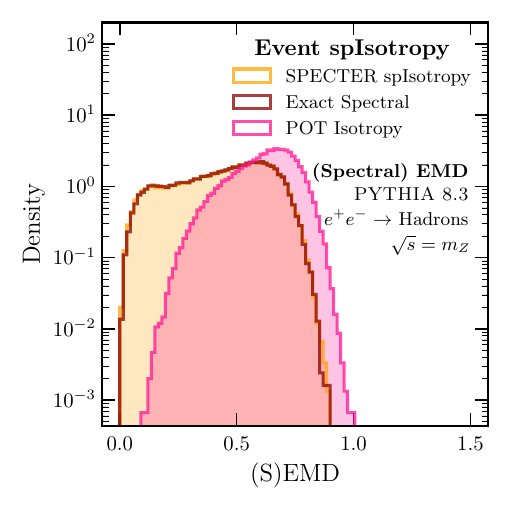}
}\label{fig:isotropy_emd}
\subfloat[]{
\includegraphics[width=0.45\textwidth]{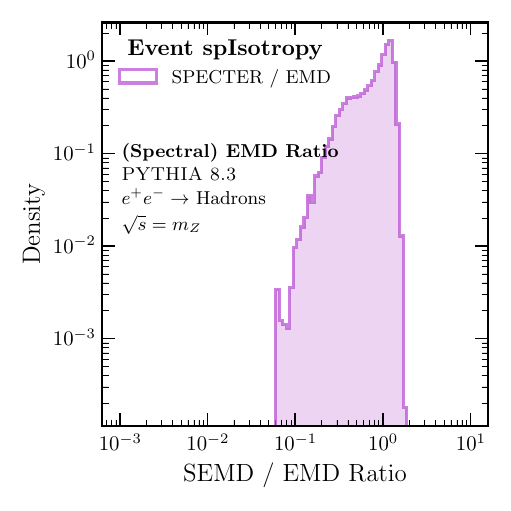}
}\label{fig:isotropy_ratio}
\caption{Distributions for the event (sp)Isotropy observable, as defined in \Sec{eventisotropy}, evaluated on jets using \Specter (orange) and the Python Optimal Transport (POT) library (pnik). The observable value is shown in (a) and the ratio of the spIsotropy to the Isotropy is shown in (b).}
\label{fig:isotropy}
\end{figure}

\begin{figure}[tpb!]
\centering
\includegraphics[width=0.45\textwidth]{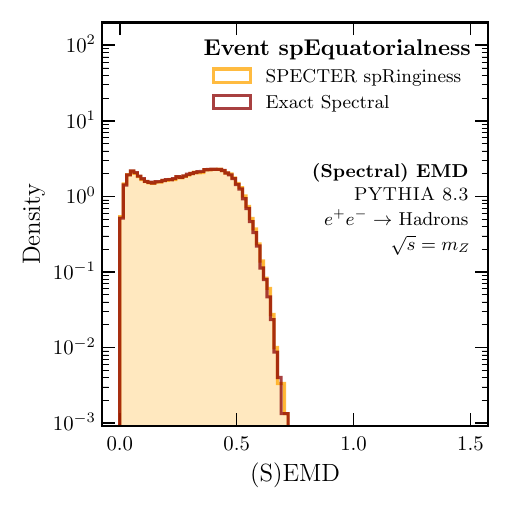}
\label{fig:event_ring_emd}
\caption{(a) Distribution for the event spEquatorialness observable, as defined in \Sec{eventringiness}, evaluated on $e^+e^-\to \text{hadrons}$ events using \Specter (orange) and exactly using \Eq{eventringobs} (dark red).}
\label{fig:event_ring}
\end{figure}

In \Fig{isotropy}, we show the spIsotropy, as computed by the  \Specter algorithm, as well as the exact closed-form expression in \Eq{spisotropyobs}.
We also show the ordinary spherical event Isotropy, as calculated using the Python Optimal Transport library~\cite{flamary2021pot}, rather than \Shaper, since \Shaper in its current form is unable to handle manifolds with local curvature.
The ordinary Isotropy is normalized such that the spIsotropy and Isotropy agree on 2-particle events.
Dijet events are not isotropic, and indeed we see that both the spIsotropy and Isotropy take large values away from zero.
As is the case with the jet shape observables, the spIsotropy tends to be strictly less than its ordinary counterpart, with the ratio $\lesssim 1$.
Unlike the event sThrust, the spIsotropy differs greatly from its ordinary counterpart, but this is to be expected: the fully isotropic event has a (nearly) highly degenerate spectral function, since it is easy to form (nearly) equilateral triangles from uniformly sampled particles, and thus we are far from the non-degenerate region probed by thrust. 
We also observe a small bump in the spIsotropy distribution near 0 -- we speculate that this may be due to degenerates caused by some additional prongy structure in the dijet events.

Finally, in \Fig{event_ring}, we show the event spEquatorialness  distribution as evaluated by \Specter, and evaluated exactly using \Eq{eventringobs}.
Here, we are unable to evaluate a corresponding observable using the ordinary EMD, as \Shaper cannot handle local curvature and the Python Optimal Transport library cannot efficiently optimize the rotational degrees of freedom of the ring over the sphere.
Interestingly, like the spIsotropy, there is also a bump in the distribution near 0, but even larger. 
These bumps appear to be a generic feature of spectral event shape observables that potentially probe interesting physics, and we leave their analysis to future work.

\section{Conclusion}\label{sec:concs}

In this paper, we introduced an efficient algorithm for evaluating the $p = 2$ SEMD.
The $p = 2$ SEMD takes on a simple closed form, which enables an exact computation between any pair of discrete events in $\O(N^2 \log N)$ time, and is to our knowledge the only exactly calculable metric on collider events.
In addition, the SEMD is \emph{fast}: using the \Specter framework, which is an efficient and highly parallelized implementation of the SEMD, it is possible to perform SEMD computations at scale, and in practical cases can even go as fast as $\O(N)$.
The SEMD is highly correlated with the ordinary EMD, especially for QCD jets, and its fast-to-evaluate, closed-form nature makes it suitable to perform precision studies of the geometry of QCD.

The speed of $\Specter$ makes previously infeasible analyses on the geometry of collider events possible. 
It is now feasible, for example, to compute distances between all ${N_{\rm Events} \choose 2}$ pairs of events in a dataset.
This, for instance, makes it easier to perform correlation dimension studies~\cite{Komiske:2022vxg}, and makes it possible to extend machine learning scaling law studies far beyond what is possible with the ordinary EMD~\cite{Batson:2023ohn}.
The idea of a metric on ``theory space'', or $\Sigma$MD, has also been proposed~\cite{Komiske:2019fks, Larkoski:2023qnv}, but to our knowledge has never been empirically studied due to being prohibitively expensive.
The ground metric of the $\Sigma$MD is itself the EMD between events, requiring ${N_{\rm Events} \choose 2}$ evaluations of the EMD. 
\Specter may be able to make the ``Spectral $\Sigma$MD''~\cite{Larkoski:2023qnv} finally tractable, which would potentially enable a novel method to probe the QCD $\beta$-function and the anomalous dimensions of other operators.

Moreover, as the SEMD is an IRC-safe metric between events, we have shown how the \Shaper prescription of \Reference{Ba:2023hix} can be carried over to define a new set of IRC-safe shape observables. 
Unlike observables defined via the ordinary EMD with \Shaper, however, the spectral observables defined via the SEMD with \Specter are far easier to evaluate due to \Specter's speed and exact metric computation, and in many cases the observables even admit exact closed-form expressions.
For instance, while the 2D ordinary event Isotropy does not seem to admit a closed form solution and is expensive to calculate, the event spIsotropy has a closed form enabling easier theoretical study, and is practically instantaneous to evaluate on huge batches of events at once.
\Specter makes the analysis of a huge class of complex jet observables far more computationally feasible, potentially higher upstream in the LHC analysis chain, and is a step closer towards being able to study complex observables at trigger level.

There are a variety of open questions and future directions not addressed in this paper.
Now that we have closed-form metrics and even closed-form shapes, an obvious next step is to attempt to compute their distributions in QCD.
While \Reference{Larkoski:2023qnv} made progress towards fixed order calculations for the $p = 1$ SEMD, is it possible to go beyond?
The EMD and SEMD seem highly correlated in QCD jets, but not other physics scenarios -- while we have speculated that this is due to the degeneracy structure of spectral space, the precise mechanism requires further explanation.
Indeed, the difference or ratio between the EMD and SEMD may itself be an interesting observable, as they behave differently depending on the context. 
Throughout this paper, we have primarily focused on balanced optimal transport, but in principle, it is possible to extend the ideas here (and the \Specter code) to accommodate unbalanced optimal transport and derived observables.
We wish to emphasize that we have only shown a small, \emph{finite} subset of possible shape observables in this paper -- the class of observables is truly infinite, and limited only by one's ability to parameterize arbitrary 2D distributions.
Finally, there exists the possibility for the \Specter code to be faster: it was developed in Python, rather than a fully compiled language, and written by a physicist rather than computer scientist -- there is almost certainly potential to be even better!

\section*{Code and Data}
\label{ref:code_data}

Version $1.0.0$ of the  \Specter framework is available at \url{https://github.com/rikab/SPECTER}.
It may also be installed via PyPi using the command $\texttt{pip install specterpy}$.
Note that a JAX installation is required for \Specter to work.
Several example and tutorial scripts for operating \Specter may be found in the subfolder \url{https://github.com/rikab/SPECTER/tree/main/examples}.
Additionally, the code used to perform all of the analyses and produce all of the plots presented in this paper may be found in \url{https://github.com/rikab/SPECTER/tree/main/studies}.

The datasets described in \Sec{datasets} are also available for use.
The \Specter package can be used to automatically download and load them -- see the examples folder.
These datasets can also be accessed independently of $\texttt{specterpy}$ using the \texttt{ParticleLoader} package, hosted at  \url{https://github.com/rikab/ParticleLoader}.

\section*{Acknowledgments}

We would like to thank Samuel Alipour-fard, Cari Cesarotti, Matt LeBlanc, and 
Melissa van Beekveld for useful and interesting discussions and comments. 

R.G. and J.T. are supported by the National Science Foundation under Cooperative Agreement PHY-2019786 (The NSF AI Institute for Artificial Intelligence and Fundamental Interactions, \url{http://iaifi.org/}), and by the U.S. DOE Office of High Energy Physics under grant number DE-SC0012567. 
J.T. is additionally supported by the Simons Foundation through
Investigator grant 929241.

\appendix

\section{More about sThrust: Chord Lengths and Energy Fractions}\label{app:thrust_chord}

The ``ordinary'' Thrust observable~\cite{Farhi:1977sg, BRANDT196457, DERUJULA1978387} is based off of the chord length between points on the celestial sphere rather than the arc length.
The sThrust also has a parameter to optimize over, the energy fraction $z$, which unlike other spectral shape observables (such as spRinginess) does not take a simple closed form.
In this appendix, we briefly describe what it would look like to use chord length as the basis for the sThrust, rather than the arc length used in \Sec{eventthrust}, and discuss the optmization of the $z$ parameter.

If we define the angular measure as
\begin{align}
\omega_{ij}^2 = \frac{ 1-\cos\theta_{ij}}{2}\,,
\end{align}
then the spectral function for a perfect, narrow dijet event is
\begin{align}
s_{\text{Thrust}^\prime}(\omega) = \left(
z^2+(1-z)^2
\right)E_\text{tot}^2\delta(\omega) + 2z(1-z)E_\text{tot}^2\delta(\omega - 1)\,,
\end{align}
on events with total energy $E_\text{tot}$ in the center-of-mass frame and energies $z E_\text{tot}$ and $(1-z)E_\text{tot}$ in each hemisphere.  With this definition of the angular measure, note that the second moment of the thrust spectral function is
\begin{align}
\int d\omega\, \omega^2\, s_{\text{Thrust}^\prime}(\omega) = 2z(1-z)E_\text{tot}^2\,,
\end{align}
and the second moment of the spectral function for a general event is
\begin{align}
\sum_{i<j\in{\cal E}}2E_iE_j\omega_{ij}^2 = \frac{1}{2}\sum_{i<j\in{\cal E}} 2E_iE_j(1-\cos\theta_{ij}) = \frac{E_\text{tot}^2}{2}\,,
\end{align}
in the center-of-mass frame.
Therefore, the SEMD between this thrust event and a general event ${\cal E}$ is
\begin{align}
\text{SEMD} &=\frac{E_\text{tot}^2}{2} 
+2z(1-z)E_\text{tot}^2\\
&\hspace{1cm}-2\sum_{\substack{n\in {\cal E}^2\\ \omega_n<\omega_{n+1}}}\omega_n \,\text{ReLU}\left(
 S^+(\omega_n)-\max\left[\left(
z^2+(1-z)^2
\right)E_\text{tot}^2, S^-(\omega_n)\right]
\right).\nonumber
\end{align}
The chord-length sThrust is then the minimum SEMD over $z$.

\begin{figure}[t!]
\centering
\subfloat[]{
\includegraphics[width=0.45\textwidth]{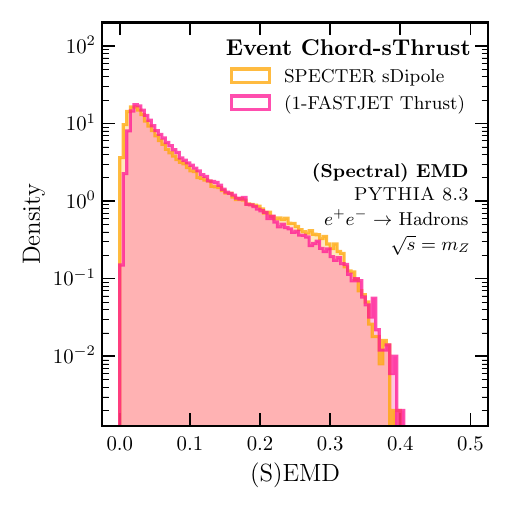}
}\label{fig:thrust_chord_emd}
\subfloat[]{
\includegraphics[width=0.45\textwidth]{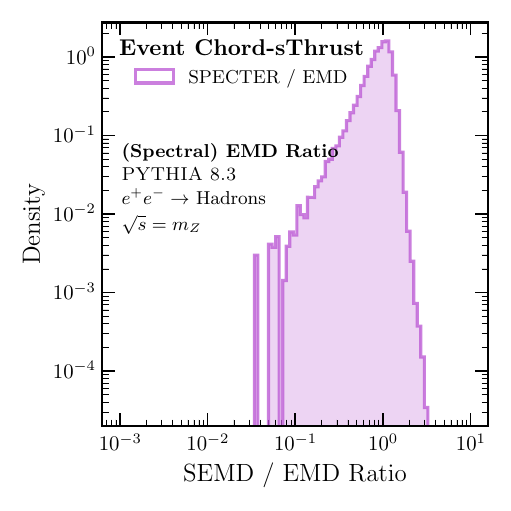}
}\label{fig:thrust_chord_ratio}
\caption{The same as \Fig{thrust}, but using the chord-based sThrust rather than the arc-length-based sThrust.}
\label{fig:thrust_chord}
\end{figure}

In \Fig{thrust_chord}, we show the distribution of this chord-based sThrust versus the ordinary Thrust. 
The distributions are practically identical, and the ratio of the two observables is sharply peaked at 1. 
Note that there is some spread around 1 -- one might expect this is due to the finite nature of the gradient descent algorithm in finding the optimal $z$ for the sThrust, but gradient descent can only ever overestimate the sThrust -- any deviations towards smaller values are due to genuine differences between the two observables. 
Nevertheless, for these back-to-back dijet events, the chord-based sThrust and Thrust are nearly identical, as expected in the collinear limit.
The differences between the two observables seen in \Fig{thrust}, then, are likely solely due to the difference in ground metric.

For both arc-length sThrust and chord-length sThrust, the value $z = \frac{1}{2}$ is significant, and empirically is almost always the minimum argument in dijet events. 
Because the spectral thrust event is symmetric under $z \leftrightarrow (1-z)$, the value $z = \frac{1}{2}$ is a local extremum.
We can check if this is a (local) maximum or a minimum with the second derivative (for the following expressions, we use the chord-length sThrust, but analogous statements hold for the arc-length sThrust with additional factors of $\pi$):
\begin{align}
\left.\frac{d^2}{dz^2}\text{SEMD}\right|_{z=1/2} =-4+8\sum_{\substack{n\in {\cal E}^2\\ \omega_n<\omega_{n+1}}}\omega_n \,\Theta\left(S^+(\omega_n)-\frac{1}{2}\right)\Theta\left(\frac{1}{2}-S^-(\omega_n)\right)\,.
\end{align}
Therefore, there is a simple test if $z = 1/2$ is the minimum.  If
\begin{align}
\sum_{\substack{n\in {\cal E}^2\\ \omega_n<\omega_{n+1}}}\omega_n \,\Theta\left(S^+(\omega_n)-\frac{1}{2}\right)\Theta\left(\frac{1}{2}-S^-(\omega_n)\right) > \frac{1}{2}\,,\label{eq:thrust_check}
\end{align}
then $z=1/2$ is a local minimum.
In fact, $z = \frac{1}{2}$ is the true \emph{global} minimum.
If \Eq{thrust_check} is true, this means that at least half the mass of the event spectral function lies to the right of $\omega = \frac{1}{2}$, which is to be optimally transported to the sThrust spectral function's peak at $\omega = 1$. 
However, this peak can only support a mass of $2z(1-z)$, which maxes out at $\frac{1}{2}$ at $z = \frac{1}{2}$, and so this is the optimal value.

On the other hand, if \Eq{thrust_check} is not true, then the minimum is displaced from $z=1/2$.  In that case, the optimal value of $z$ then corresponds to the amount of squared energy transported from the region where $\omega<1/2$ to $\omega = 0$, as the solution of
\begin{align}
z^2 +(1-z)^2 = \frac{1}{E^2_\text{tot}}\int_0^{1/2}d\omega\, s(\omega)\,.
\end{align}
The precise value of $z$ actually isn't that important, because the optimal transport plan for the SEMD is to move all energy at less than $\omega=1/2$ to $\omega = 0$, and to move all energy above $\omega = 1/2$ to $\omega = 1$.

\section{Closed-Form Expressions for Spectral Observables}

Many of the observables in \Sec{specshapes} admit exact closed-form expressions.
In this appendix, we go over the derivations of these expressions.

\subsection{Closed-Form Jet spRinginess}\label{app:closed_form_jet_springiness}

For a uniform energy distribution in the shape of a circle of radius $R$ in a jet with total energy $E_\text{tot}$, the cumulative spectral function is
\begin{align}
S_\text{Jet Ring}(\omega) = \frac{2E_\text{tot}^2}{\pi}\,\sin^{-1}\left(\frac{\omega}{2R}\right)\,.
\end{align}
The inverse cumulative spectral function is then
\begin{align}
S_\text{Jet Ring}^{-1}(E^2) = 2R\, \sin\left(
\frac{\pi}{2}\frac{E^2}{E_\text{tot}^2}
\right)\,.
\end{align}
The integral of the square of this is 
\begin{align}
\int_0^{E_\text{tot}^2} dE^2\, \left(
S_\text{Jet Ring}^{-1}(E^2)
\right)^2 = 2E_\text{tot}^2 R^2\,.
\end{align}
We also need the integral 
\begin{align}
\int dE^2\, S^{-1}(E^2)S^{-1}_\text{Jet Ring}(E^2) &= 2R\sum_{\substack{n\in{\cal E}^2\\\omega_n<\omega_{n+1}}}\omega_n\int_{S^-(\omega_n)}^{S^+(\omega_n)} dE^2\, \sin\left(\frac{\pi}{2} \frac{E^2}{E_\text{tot}^2}\right)\\
&\hspace{-2cm}=\frac{4R}{\pi}E_\text{tot}^2\sum_{\substack{n\in{\cal E}^2\\\omega_n<\omega_{n+1}}}\omega_n\left[
\cos\left(
\frac{\pi}{2E_\text{tot}^2}S^-(\omega_n)
\right)-\cos\left(
\frac{\pi}{2E_\text{tot}^2}
S^+(\omega_n)
\right)
\right]\nonumber\,.
\end{align}

With these results, we note that the SEMD between a jet and the uniform ring configuration is quadratic and convex in the radius parameter $R$.  As such, there is a unique, global minimum that can be determined and established as the optimal value of the radius.  This optimal radius $R_{\text{opt}}(s_\E)$ is
\begin{align}
R_\text{opt} = \frac{2}{\pi}\sum_{\substack{n\in{\cal E}^2\\\omega_n<\omega_{n+1}}}\omega_n\left[
\cos\left(
\frac{\pi}{2E_\text{tot}^2}S^-(\omega_n)
\right)-\cos\left(
\frac{\pi}{2E_\text{tot}^2}
S^+(\omega_n)
\right)
\right]\,.
\end{align}
These results can then be used to construct the observable of \Eq{semd_jet_ring}.

\subsection{Closed-Form spLineliness}\label{app:closed_form_splineliness}

For a continuous energy distribution arranged along a line of length $L$ in a jet, its cumulative spectral function is
\begin{align}
S_\text{line}(\omega) =\left( \frac{2\omega}{L}-\frac{\omega^2}{L^2}\right)E_\text{tot}^2\,.
\end{align}
The inverse cumulative spectral function is then
\begin{align}
S_\text{line}^{-1}(E^2) = \left(
1-\sqrt{1-\frac{E^2}{E_\text{tot}^2}}
\right)L\,.
\end{align}
Its squared integral is then
\begin{align}
\int dE^2\, S_\text{line}^{-1}(E^2)^2 = \frac{L^2}{6}\, E_\text{tot}^2\,.
\end{align}
Next, we need to integrate this inverse cumulative spectral function against the inverse cumulative spectral function for the event of interest.  We have
\begin{align}
\int dE^2\, S^{-1}(E^2)S^{-1}_\text{line}(E^2) &= L\sum_{\substack{n\in{\cal E}^2\\\omega_n<\omega_{n+1}}}\omega_n\int_{S^-(\omega_n)}^{S^+(\omega_n)} dE^2\,\left(
1-\sqrt{1-\frac{E^2}{E_\text{tot}^2}}
\right)\\
&=L\sum_{\substack{n\in{\cal E}^2\\\omega_n<\omega_{n+1}}}\omega_n\left[
S^+(\omega_n)+\frac{2}{3E_\text{tot}}\left(
E_\text{tot}^2-S^+(\omega_n)
\right)^{3/2}\right.\nonumber\\
&
\hspace{4cm}\left.-\,S^-(\omega_n)-\frac{2}{3E_\text{tot}}\left(
E_\text{tot}^2-S^-(\omega_n)
\right)^{3/2}
\right]\,.\nonumber
\end{align}
As we observed with the jet spRinginess, the SEMD between the uniform line configuration and a jet is quadratic and convex in the length parameter $L$, and so it can be optimized by setting it equal to its unique minimum.  These results can then be used to construct the observable of \Eq{jetlineobs}.

\subsection{Closed-Form spIsotropy}\label{app:closed_form_spisotropy}

Consider a continuous energy distribution on the celestial sphere that is perfectly isotropic.  The spectral function for this energy distribution can be calculated by fixing a reference particle to be located at the north pole, and the other particle at some point with polar angle $\theta$ and azimuthal angle $\phi$.  The spectral function then follows from
\begin{align}
s_\text{iso}(\omega) = \frac{E_\text{tot}^2}{4\pi}\int_0^{2\pi} d\phi\int_0^\pi\sin\theta\, d\theta\, \delta(\omega - \theta) = \frac{\sin\omega}{2}\, E_\text{tot}^2\,,
\end{align}
for a total energy $E_\text{tot}$.  The cumulative spectral function of angular distances $\omega$ on the sphere is then
\begin{align}
S_\text{iso}(\omega) =\frac{1-\cos\omega}{2} E_\text{tot}^2\,.
\end{align}
The inverse cumulative spectral function is then
\begin{align}
S_\text{iso}^{-1}(E^2) = \cos^{-1}\left(
1-2\frac{E^2}{E_\text{tot}^2}
\right)\,.
\end{align}
The integral of its square is
\begin{align}
\int_0^{E_\text{tot}^2}dE^2\, S_\text{iso}^{-1}(E^2)^2 = \frac{\pi^2-4}{2}\,E_\text{tot}^2\,.
\end{align}

Next, we need to integrate this inverse cumulative spectral function against the inverse cumulative spectral function for the event of interest.  We have
\begin{align}
\int dE^2\, S^{-1}(E^2)S^{-1}_\text{iso}(E^2) &= \sum_{\substack{n\in{\cal E}^2\\\omega_n<\omega_{n+1}}}\omega_n\int_{S^-(\omega_n)}^{S^+(\omega_n)} dE^2\,\cos^{-1}\left(
1-2\frac{E^2}{E_\text{tot}^2}
\right)\\
&\hspace{-4cm}=\sum_{\substack{n\in{\cal E}^2\\\omega_n<\omega_{n+1}}}\omega_n\left[
\sqrt{S^+(\omega_n)}\sqrt{E_\text{tot}^2-S^+(\omega_n)}+S^+(\omega_n)\,\cos^{-1}\left(
1-2\frac{S^+(\omega_n)}{E_\text{tot}^2}
\right)-E_\text{tot}^2\,\sin^{-1}\left(
\frac{\sqrt{S^+(\omega_n)}}{E_\text{tot}}
\right)
\nonumber\right.\\
&\hspace{-3cm}\left.-\sqrt{S^-(\omega_n)}\sqrt{E_\text{tot}^2-S^-(\omega_n)}-S^-(\omega_n)\,\cos^{-1}\left(
1-2\frac{S^-(\omega_n)}{E_\text{tot}^2}
\right)+E_\text{tot}^2\,\sin^{-1}\left(
\frac{\sqrt{S^-(\omega_n)}}{E_\text{tot}}
\right)
\right].\nonumber
\end{align}
These results can then be used to construct the expression for the observable of \Eq{spisotropyobs}.

\subsection{Closed-Form Event spEquatorialness}\label{app:closed_form_event_springiness}

For continuous energy distribution on the equator of the celestial sphere, the cumulative spectral function is
\begin{align}
S_\text{ring}(\omega) = \frac{\omega}{\pi}\,E_\text{tot}^2\,.
\end{align}
The inverse cumulative spectral function is then
\begin{align}
S_\text{ring}^{-1}(E^2) = \pi\, \frac{E^2}{E_\text{tot}^2}\,.
\end{align}
The integral of the square of this inverse cumulative spectral function is
\begin{align}
\int_0^{E_\text{tot}^2}dE^2\, S_\text{ring}^{-1}(E^2)^2 =\frac{\pi^2}{3}\,E_\text{tot}^2\,.
\end{align}

Next, we need to integrate this inverse cumulative spectral function against the inverse cumulative spectral function for the event of interest.  We have
\begin{align}
\int dE^2\, S^{-1}(E^2)S^{-1}_\text{ring}(E^2) &= \sum_{\substack{n\in{\cal E}^2\\\omega_n<\omega_{n+1}}}\omega_n\int_{S^-(\omega_n)}^{S^+(\omega_n)} dE^2\, \pi\, \frac{E^2}{E_\text{tot}^2}\\
&=\frac{\pi}{2}\frac{1}{E_\text{tot}^2}\sum_{\substack{n\in{\cal E}^2\\\omega_n<\omega_{n+1}}}\omega_n\left[
\left(
S^+(\omega_n)
\right)^2-\left(
S^-(\omega_n)
\right)^2
\right]\nonumber\,.
\end{align}
This can then be used to construct the observable of \Eq{eventringobs}.

\bibliographystyle{JHEP}
\bibliography{refs, jet_refs}

\end{document}